\documentclass[letterpaper,twocolumn,10pt]{article}

\usepackage{tikz}
\usepackage{amsmath}

\usepackage{cite}
\usepackage{amsmath,amssymb,amsfonts,amsthm}
\usepackage{usenix}
\usepackage{algorithm}
\usepackage{graphicx}
\usepackage{balance}
\usepackage{textcomp}
\usepackage{xcolor}
\usepackage{longtable}
\usepackage{multirow}
\usepackage{float}
\usepackage[utf8]{inputenc}
\usepackage{pifont,textcomp}
\usepackage{booktabs, tabularx, array, ragged2e}
\usepackage{comment}
\usepackage{textgreek} 
\usepackage{algpseudocode}
\usepackage[inline]{enumitem}
\usepackage{etoolbox}
\appto\UrlBreaks{\do\-\do\.\do\/\do\_}
\usepackage{filecontents}
\usepackage{cleveref}
\usepackage{booktabs} 
\usepackage{array}    
\usepackage{caption} 
\usepackage{siunitx}  
\usepackage{subcaption} 
\usepackage{xspace}
\usepackage{tcolorbox}  
\usepackage{placeins}  
\usepackage{bm}
\tcbuselibrary{breakable}
\usepackage{xcolor}

\newcommand{\atlas}{ATLAS\xspace}
\newcommand{\airtag}{\textsc{AirTag}\xspace}
\newcommand{\tool}{\textsc{Ananke}\xspace}

\begin{document}

\title{\Large \bf  An Automated Attack Investigation Approach Leveraging Threat-Knowledge-Augmented Large Language Models }

\author{
Rujie Dai$^{1,2,*}$,
Peizhuo Lv$^{3,*,\dag}$,
Yujiang Gui$^{4}$,
Qiujian Lv$^{1,2}$,
Yuanyuan Qiao$^{5}$,
Yan Wang$^{1,2,\dag}$,\\
Degang Sun$^{6}$,
Weiqing Huang$^{1,2}$,
Yingjiu Li$^{7}$,
and XiaoFeng Wang$^{3}$\\[0.3em]
\textit{$^{1}$Institute of Information Engineering, Chinese Academy of Sciences, China}\\
\textit{$^{2}$University of Chinese Academy of Sciences, China}\\
\textit{$^{3}$Nanyang Technological University, Singapore}\\
\textit{$^{4}$University of New South Wales, Australia}\\
\textit{$^{5}$Beijing University of Posts and Telecommunications, China}\\
\textit{$^{6}$Computer Network Information Center, Chinese Academy of Sciences, China}\\
\textit{$^{7}$University of Oregon, USA}\\[0.6em]
{\rmfamily \{dairujie2024, lvqiujian, wangyan, huangweiqing\}@iie.ac.cn}\\
{\rmfamily peizhuo.lyu@ntu.edu.sg, yujiang.gui@unsw.edu.au, yyqiao@bupt.edu.cn}\\
{\rmfamily dgsun@cnic.cn, yingjiul@uoregon.edu, xiaofeng.wang@ntu.edu.sg}\\[0.6em]
$^*$Co-first authors \quad $^\dag$Corresponding authors
}

\maketitle

\begin{abstract}

Advanced Persistent Threats (APTs) are prolonged, stealthy intrusions by skilled adversaries that compromise high-value systems to steal data or disrupt operations. Reconstructing complete attack chains from massive, heterogeneous logs is essential for effective attack investigation, yet existing methods suffer from poor platform generality, limited generalization to evolving tactics, and an inability to produce analyst-ready reports. Large Language Models (LLMs) offer strong semantic understanding and summarization capabilities, but in this domain they struggle to capture the long-range, cross-log dependencies critical for accurate reconstruction.

To solve these problems, we present an LLM-empowered attack investigation framework augmented with a dynamically adaptable Kill-Chain-aligned threat knowledge base. We organizes attack-relevant behaviors into stage-aware knowledge units enriched with semantic annotations, enabling the LLM to iteratively retrieve relevant intelligence, perform causal reasoning, and progressively expand the investigation context. This process reconstructs multi-phase attack scenarios and generates coherent, human-readable investigation reports. Evaluated on 15 attack scenarios spanning single-host and multi-host environments across Windows and Linux (over 4.3M log events, 7.2 GB of data), the system achieves an average True Positive Rate (TPR) of 97.1\% and an average False Positive Rate (FPR) of 0.2\%, significantly outperforming the SOTA method \atlas, which achieves an average TPR of 79.2\% and an average FPR of 29.1\%.

\end{abstract}

\section{Introduction}

Advanced Persistent Threats (APTs) are deliberate, highly targeted intrusions in which skilled adversaries compromise organizations, remain covert for extended periods, and cause severe damage~\cite{damage:mitreSolarWindsCompromise,damage:finance_solarwinds_techrepublicCybersecurityStudy,soloarwinds:cisaActiveExploitation,damage:kasperskyFinancialAttacks,damage:revealAPT28Cyber}. Timely and comprehensive reconstruction of attack chains is therefore essential to support incident reporting, improve defense strategies, and strengthen collective resilience~\cite{invet_import:theguardianSellafieldApologises,invet_import:upguardCyberIncident}.
Starting with security alerts~\cite{attackIvestigation:cloudopticsThreatHunting}, attack investigation reconstructs how intrusions unfolded~\cite{attack_in_fromlogs:ji2017rain,attack_in_fromlogs:zhu2011attack,attack_in_fromlogs:lee2013high} by analyzing large volumes of noisy logs to reveal adversary \textit{tactics, techniques, and procedures} (TTPs~\cite{ttps:cymulateTacticsTechniques}) and produce human-readable reports~\cite{CTI:crowdstrikeWhatCyber,using_report:perry2019no}, thereby supporting rapid incident response, defense planning, and threat intelligence accumulation.

Early research employed heuristics and handcrafted rules to trace malicious behavior in system logs~\cite{DBLP:conf/ndss/MaZX16,provance_graphmatch:milajerdi2019poirot,relatedwork_rule_based:milajerdi2019holmes,relatedwork_rule_based:hossain2017sleuth}.
To reduce manual effort, recent research~\cite{airtag_2023,atlas_2021,shen2019attack2vec,deeplog:du2017deeplog,provance:han2020unicorn,relatedwork_static_based:hassan2019nodoze,kairos:cheng2024kairos,yang2023prographer} has explored deep learning for intrusion and anomaly detection and analysis. These methods learn either sequence-level patterns ~\cite{airtag_2023,atlas_2021,shen2019attack2vec} or feature-level representations ~\cite{deeplog:du2017deeplog,provance:han2020unicorn} from logs to identify suspicious behaviors. Other studies have applied graph-based learning to capture structural relationships that may indicate malicious activity ~\cite{relatedwork_static_based:hassan2019nodoze,kairos:cheng2024kairos,yang2023prographer}.
Despite these advances, existing approaches face fundamental limitations. Handcrafted rule-based systems~\cite{provance_graphmatch:milajerdi2019poirot,relatedwork_rule_based:milajerdi2019holmes,relatedwork_rule_based:hossain2017sleuth} are labor-intensive and fail to adapt to evolving attack patterns, while learning-based methods~\cite{yang2023prographer,jia2024magic,relatedwork_rule_graph_based:dong2023distdet} depend on historical data and exhibit poor generalizability in response to novel attacks and diverse benign behaviors. For example, \textsc{DistDet}~\cite{relatedwork_rule_graph_based:dong2023distdet} reports poor generalization across benign behaviors. Moreover, adapting to new attack types requires retraining or fine-tuning, which is costly, risks catastrophic forgetting~\cite{catastrophic:kemker2018measuring}, and delays deployment. Another challenge is cross-platform inconsistency: system logs are tied to platform-specific mechanisms (e.g., Windows ETW vs. Linux Audit), and approaches such as \atlas~\cite{atlas_2021} must build separate vocabularies for each OS, causing performance degradation across platforms~\cite{logs_feature:wu2023effectiveness,logs_feature:jilcha2025temporal}. Finally, many systems~\cite{airtag_2023,atlas_2021,relatedwork_rule_based:hossain2017sleuth} output raw snippets rather than coherent narratives, forcing analysts to assemble attack stories manually.

Trained on massive and diverse corpora, LLMs~\cite{llm:zhao2023survey,llm_reasoning:wei2022chain} are capable of understanding heterogeneous logs, capturing contextual dependencies, and generating analyst-readable summaries, which offer a promising alternative to perform attack investigation. However, they lack structured threat knowledge~\cite{llm_domain:wang2023survey} and struggle to process long-duration, high-volume logs within their limited context windows, leading to fragmented reasoning and incomplete narratives.

To overcome these limitations, we propose \tool, the first LLM-empowered attack investigation framework grounded in Retrieval-Augmented Generation (RAG) ~\cite{relatedwork_rag:lewis2020retrieval} with a dynamically adaptable threat knowledge base aligned to the Cyber Kill Chain~\cite{lockheedmartinCyberKill}.
Since attacks unfold through sequential phases with causal dependencies, the Kill Chain provides a natural structure to guide LLM reasoning.
\tool automatically constructs and enriches a phase-specific knowledge base from public attack logs by leveraging LLMs’ foundational understanding of adversarial tactics, reducing expert effort while enabling training-free adaptation to evolving threats. This design eliminates costly retraining, mitigates catastrophic forgetting, and augments LLM reasoning by retrieving relevant threat knowledge as contextual guidance.
Furthermore, since LLMs cannot process massive logs in a single pass or maintain causal continuity alone, \tool employs a graph-based expansion with iterative reasoning and a cache to preserve context across iterations. 
Together, these mechanisms allow \tool to progressively reconstruct complete attack chains and generate coherent, analyst-ready forensic reports in a scalable and explainable manner.

\tool has been evaluated on both public and simulated datasets, including \atlas dataset~\cite{atlas_2021}, \textsc{DepImpact} dataset ~\cite{relatedwork_rule_based:fang2022back}, as well as two simulated shellcode injection scenarios (EB-P1 and EB-P2). These datasets collectively comprise over 4.3 million log events totaling 7.2 GB. Evaluation results show that \tool achieves an average TPR of 97.1\% and an FPR of only 0.2\% across diverse attack settings, including those where malicious code is injected into benign processes. Importantly, without retraining, \tool maintains strong performance against novel attacks (e.g., 80.5\% TPR and 0.1\% FPR on EB-P1, compared to ATLAS with 21.7\% TPR and 2.0\% FPR) and generalizes to Linux datasets (90.0\% TPR and 0.9\% FPR, whereas ATLAS achieves 69.6\% TPR and 56.3\% FPR) without platform-specific feature engineering. All experiments produce coherent, analyst-readable reports, underscoring the practicality of the approach.

\noindent\textbf{Contributions.} We summarize our contributions as below:
\begin{itemize}[left=0em]
\item We propose \tool, the first framework to leverage LLMs for automated attack investigation, reconstructing attack scenarios from massive heterogeneous logs and generating coherent, analyst-ready reports.

\item \tool organizes a Cyber Kill Chain-based threat knowledge base and retrieves relevant adversarial knowledge in a RAG manner to augment LLM reasoning, and employs graph-based expansion to progressively extend attack context, enabling step-by-step reconstruction of complete attack scenarios without feature engineering or retraining.

\item We evaluate \tool on public and custom datasets, demonstrating strong accuracy, adaptability, and generalization across diverse attacks and platforms. To foster reproducibility and future research, we release our implementation, datasets, and a public demo~\cite{code:4openAnonymousGithub}.
\end{itemize}

\section{Background}\label{sec:Background}

\subsection{Provenance Data and Properties}
\label{sec:bg:provenance_data}

Modern operating systems provide monitoring frameworks and tools such as Event Tracing for Windows (ETW) \cite{etw_microsoftEventTracing}, Sysmon \cite{sysmon；microsoftSysmonSysinternals}, Linux Audit \cite{linux_audit_redhatChapterxA07xA0SystemAuditing}, and Sysdig \cite{sysdigSecurityTools}, which record fine-grained system activities and form a fundamental data source for forensic analysis.

\begin{table}[htbp]
\renewcommand{\arraystretch}{0.9}
\centering
\caption{Examples of System Entities and Their Activities}
\label{tab:triplet_definition}
\resizebox{\linewidth}{!}{
\begin{tabular}{ll}
\toprule
\textbf{Entity Type\kern 4em} & \textbf{Example Actions\kern 3em} \\
\midrule
Process & Execute, Fork, Bind, Close, etc. \\
File    & Read, Write, Delete, Rename, etc. \\
Socket  & Send, Receive, Listen, Accept, etc. \\
Website & Resolve, Request, etc. \\
IP Address & Connect, Disconnect, etc. \\
\bottomrule
\end{tabular}
}
\end{table}

Audit logs can be abstracted as n \emph{subject–action–object} triples (\Cref{tab:triplet_definition}) and can be defined as structured event triples \(e = \langle s, a, o \rangle\), where $s$ and $o$ represent subject and object entities (e.g., processes, files), and $a$ denotes the action (e.g., read, write). These events are composed into a directed provenance graph \(G = (V, E)\), with vertices $V$ denoting system entities and edges $E$ representing timestamped interactions. For instance, a process (subject) may write (action) to a file (object).

This abstraction highlights three properties that are critical to forensic analysis:
\begin{enumerate}[leftmargin=*,topsep=2pt,itemsep=1pt]
  \item \emph{Causality.} Actor–action links event dependencies to support provenance tracing.
  \item \emph{Temporality.} High-resolution timestamps enable global ordering and cross-host correlation.
  \item \emph{Semantics.} Human-readable fields reveal intent, aiding obfuscation detection and policy enforcement.
\end{enumerate}
By exposing these properties, the structured triples and the resulting provenance graph establish a unified representation that underpins downstream security tasks such as anomaly detection and attack investigation.

\subsection{Kill Chain}
\label{sec:bg:killchain}
The  Kill Chain, introduced by Lockheed Martin, is a widely adopted framework for modeling the progression of targeted cyber attacks~\cite{lockheedmartinCyberKill}. It divides an attack into seven sequential phases: \textit{Reconnaissance, Weaponization, Delivery, Exploitation, Installation, Command and Control}, and \textit{Actions on Objectives}. This structure allows defenders to systematically interpret attacker behavior, correlate events across heterogeneous logs, and apply phase-specific detection and response strategies. By offering a unified language for describing attack workflows, the Kill Chain forms a foundational basis for threat analysis, intrusion detection, and forensic investigation.

\subsection{Attack Investigation}
\label{sec:bg:investigation}
Intrusion Detection Systems (IDS) generate alerts for potentially malicious activity, typically serving as entry points for attack investigations. Given a triggered alert, the goal is to reconstruct the full attack path by analyzing causal dependencies in system audit logs.

\begin{figure}[htbp]
  \centering
  \includegraphics[width=0.5\textwidth,trim=0.1cm 1.0cm 0cm 0cm, clip]{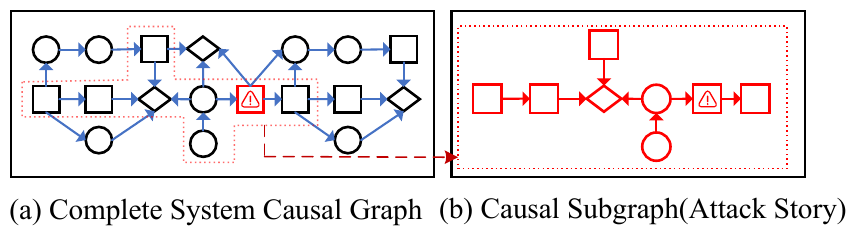}
  \begin{subfigure}{0.24\textwidth}
      \caption{Complete system causal graph}
  \end{subfigure}%
  \begin{subfigure}{0.26\textwidth}
      \caption{Causal subgraph}
  \end{subfigure}%
  \caption{Extraction of an attack causal subgraph from the complete system causal graph to reveal the attack story.}
  \label{fig:attack_investigation}
\end{figure}

We follow the same representation of audit logs as structured event triples and provenance graph construction described in Section~\ref{sec:bg:provenance_data}. 
Given an alert entity node $v_{\text{alert}} \in V$, the investigation aims to extract a subgraph $G_{\text{attack}} \subseteq G$ that captures the relevant causal context:
\begin{equation}
G_{\text{attack}}\!=\!\mathrm{TraceForward}(v_{\text{alert}}) \cup \mathrm{TraceBackward}(v_{\text{alert}})
\label{eq:attack_subgraph}
\end{equation}
This involves bounded forward and backward graph traversals that uncover temporal and logical event relationships. Manual correlation is time-consuming and error-prone, motivating automated approaches to scale investigation over heterogeneous log sources (as illustrated in Figure~\ref{fig:attack_investigation}).

\subsection{Retrieval-Augmented Generation}
\label{sec:bg:rag}

To avoid fine-tuning costs, Retrieval-Augmented Generation (RAG) enables LLMs to access external knowledge dynamically. A RAG system operates in two phases:

\smallskip
\textit{Knowledge Retrieval.} Given a query $Q$ and knowledge base $D = \{d_1, \dots, d_N\}$, the system first computes the semantic similarity between the query embedding $e(Q)$ and each document embedding $e(d_i)$ in the document knowledge base. The relevance score between $e(Q)$ and $e(d_i)$ is computed using a similarity function as follows:
\[
\mathit{sim}(e(Q), e(d_i)) = \frac{e(Q) \cdot e(d_i)}{\|e(Q)\| \|e(d_i)\|}, \quad \forall e(d_i) \in D
\]
The top-$k$ most relevant documents $d_i$ with the highest similarity scores are selected to form the retrieval result:
\[
\{d_1, d_2, \ldots, d_k\} = \operatorname{top-}k \bigl(\mathit{sim}(e(Q), e(d_i))\bigr),\quad d_i \in D
\]

\textit{Augmented Generation.} The final answer is generated by LLM through contextual aggregation. Formally, the answer generation process is defined as:
\[
\hat{a}(Q) = \mathrm{LLM}\bigl(Q, \{ d_1, \ldots, d_k \} \bigr)
\label{eq:answer_gen}
\]
Here, \( Q \) denotes the current query, and \( \{ d_1, \ldots, d_k \} \) represents the retrieved context units. 
The large language model \( \mathrm{LLM}(\cdot) \) takes both the query and the retrieved context as input and produces the final response \( \hat{a}(Q) \).

RAG enables training-free adaptation to new domains, improving answer relevance and supporting domain-aware reasoning without modifying model parameters. Following the retrieve-then-generate paradigm, its effectiveness depends directly on the quality of retrieval.

\begin{figure}[h]
  \centering
  \includegraphics[width=0.4\textwidth,trim=0.4cm 0.1cm 0cm 0cm, clip]{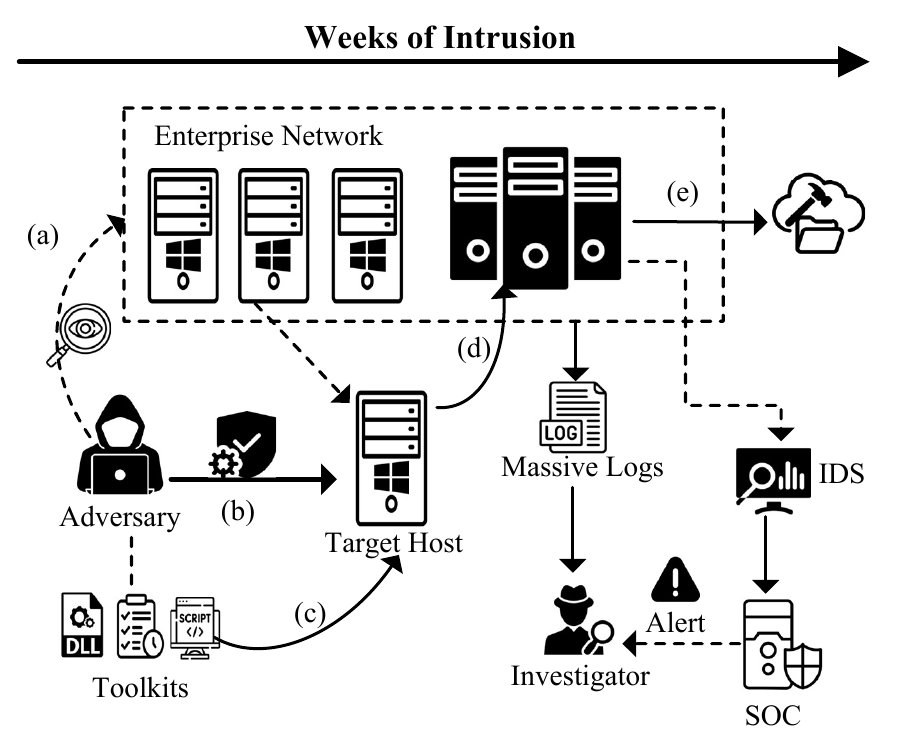}
  \caption{An intrusion leveraging the EternalBlue vulnerability (\texttt{CVE-2017-0144}). (a) An adversary scans exposed SMB endpoints, (b) exploits a vulnerable Windows server, (c) establishes persistence, (d) moves laterally across heterogeneous hosts, and (e) exfiltrates sensitive data to cloud storage.}
\label{fig:motivating_example}
\end{figure}

\section{Problem Statement}

\label{sec:Motivation}

\subsection{A Motivating Example}

We illustrate the importance of attack investigation with an attack exploiting the EternalBlue vulnerability (\texttt{CVE-2017-}\linebreak\texttt{0144}), a Windows SMB flaw originally developed by the NSA and later leaked by the Shadow Brokers, which fueled large-scale campaigns such as WannaCry and NotPetya. Consider a multinational enterprise with a hybrid infrastructure where legacy Windows servers coexist with Linux-based services. Its Security Operations Center (SOC) deploys Intrusion Detection Systems (IDS) to monitor network flows and host activities, supported by security analysts and forensic tools.

The adversary first scans exposed services to identify vulnerable SMB endpoints (Figure~\ref{fig:motivating_example} (a)). Upon discovering an unpatched Windows server, the attacker exploits EternalBlue to gain SYSTEM-level privileges (b). With this foothold, persistence is established via scheduled tasks, DLL hijacking, or credential theft (c). Using stolen credentials, the attacker moves laterally across the enterprise (d), exploiting SMB sessions on additional Windows hosts and tunneling through RDP into Linux management servers. Finally, sensitive enterprise data is exfiltrated to external cloud storage (e).

During this process, the SOC only observes isolated anomalies (e.g., suspicious process activities) without visibility into the full attack chain. The intrusion persists for weeks, generating gigabytes of heterogeneous logs. Dynamic evasion strategies obscure traces, while delayed retraining of detection models further reduces accuracy. Even after filtering, analysts must spend weeks manually correlating timestamps and host activities before reconstructing the complete intrusion.

\subsection{Threat Model}
\label{threat_model}
Similar to prior studies~\cite{provance:han2020unicorn,relatedwork_rule_based:milajerdi2019holmes,relatedwork_rule:milajerdi2019poirot,kairos:cheng2024kairos,atlas_2021,zengy2022shadewatcher,jia2024magic}, we consider an adversary who obtains initial execution by exploiting software vulnerabilities, delivering malicious documents or links, or invoking remote-code-execution channels. After establishing a foothold, the attacker may pursue persistence, privilege escalation, data exfiltration, and lateral movement.

We assume that the defender is a cyber attack investigator, whose work targets \textit{post-compromise analysi}s and does not attempt to block intrusions in real time. The trusted computing base (TCB) consists of the system hardware, operating system, audit framework, and our analysis code, consistent with assumptions made in prior forensic and provenance-based intrusion detection systems~\cite{kairos:cheng2024kairos,jia2024magic}. We assume that existing system hardening techniques mitigate audit framework compromise and exclude kernel-level attacks from our threat model. Furthermore, we consider that existing logging infrastructures~\cite{threat_model_tcb:paccagnella2020custos,threat_model_tcb:paccagnella2020logging}, implemented through append-only storage or hardware-assisted secure logging, are capable of ensuring tamper-evidence. Under this assumption, the collected log corpus is regarded as an accurate and complete record of system behavior during the investigation period.

In addition, the defender is also assumed to have access to raw logs containing known attacks from open-source threat intelligence platforms~\cite{attack_log_cti:crowdstrikeThreatIntelligence,mitre_ttps,attack_log_cti:levelblueLevelBlueLabs}, sandbox environments\cite{attack_log_sandbox:anyANYRUNInteractive}, security vendors\cite{attack_log_security_vendor:talosintelligenceCiscoTalos}, or publicly available attack datasets\cite{attack_log_public_datasets:githubGitHubFiveDirectionsOpTCdata,attack_log_public_datasets:githubGitHubShramosAwesomeCybersecurityDatasets,attack_log_public_datasets:lanlDataSets}. As attacks evolve, updated open-source intelligence and public datasets enable the defender to continually enrich the threat knowledge base. Based on these logs, the defender can construct a dynamically adaptable threat knowledge base that facilitates investigation.

\begin{figure}[h]
    \centering
    \begin{subfigure}[]{0.24\textwidth}  
        \centering
        \includegraphics[width=\linewidth]{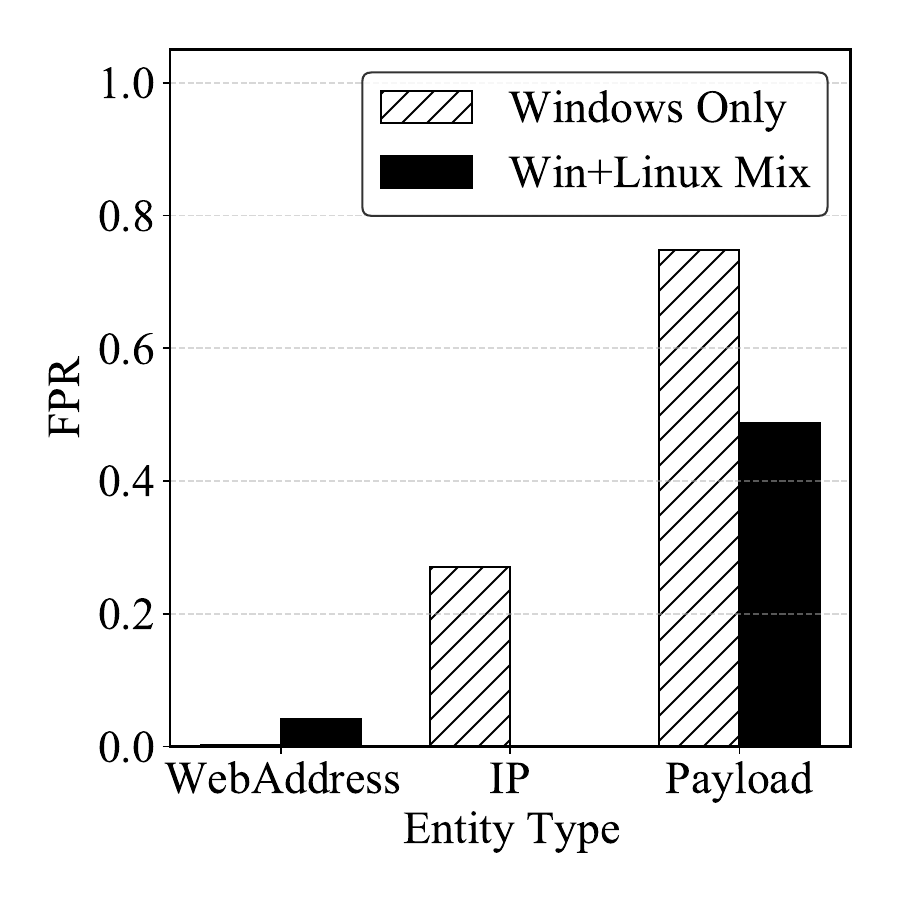}
        \caption{TPR on mixed training data.}
        \label{fig:FPR_Comparison_Win_vs_Mix_bw}
    \end{subfigure}%
    \begin{subfigure}[]{0.24\textwidth}  
        \centering
        \includegraphics[width=\linewidth]{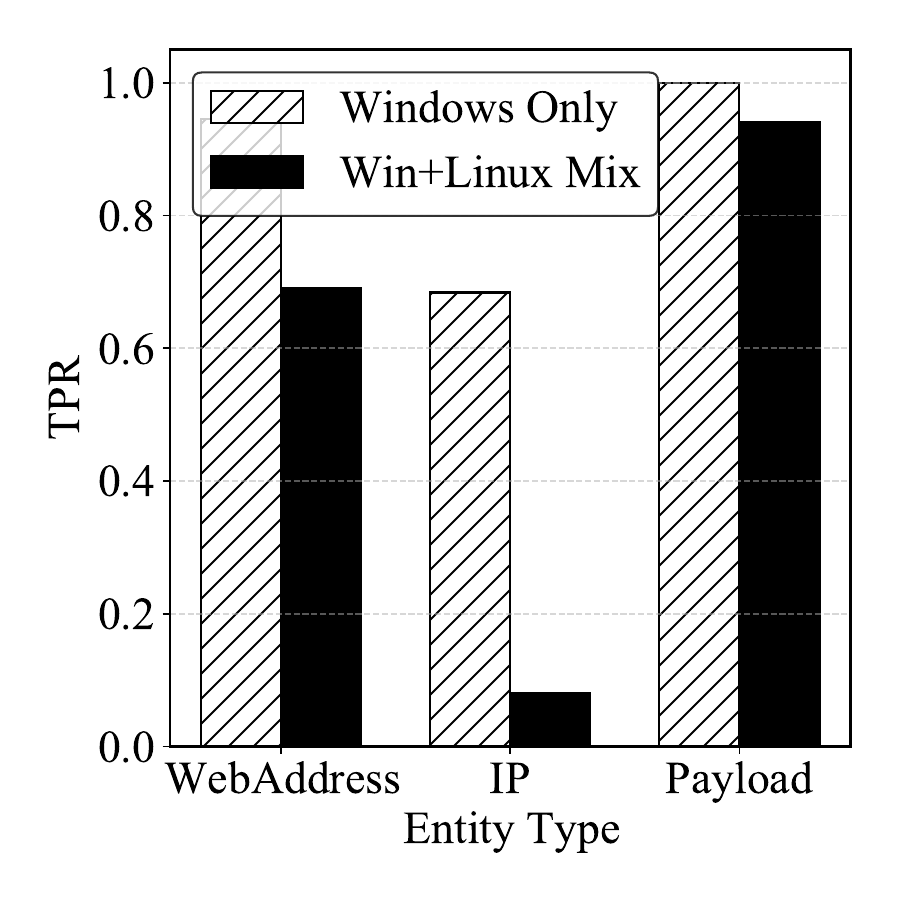}
        \caption{FPR on mixed training data.}
        \label{fig:TPR_Comparison_Win_vs_Mix}
    \end{subfigure}%
    
    \caption{Impact of mixed training data.}
    \label{fig:Impact_mixed_training_data}
\end{figure}

\subsection{Technical Challenges}
\label{sec:challenges}
APT attackers continually evolve their tactics, tools, and procedures across time and platforms.
Many learning-based methods~\cite{yang2023prographer,atlas_2021,relatedwork_rule_based:fang2022back,relatedwork_rule_graph_based:dong2023distdet}, trained on past attack patterns, require costly retraining or fine-tuning to adapt to new attacks and risk forgetting previous knowledge~\cite{catastrophic:kemker2018measuring} due to limited generalization.
For example, an adversary may switch from exploiting a Flash vulnerability to exploiting an SMB service vulnerability to inject malicious code into normal processes and gain control. Such drastic changes can cause models to fail to adapt to new attacks, forcing costly retraining and delaying deployment.
The vast volume of logs makes attack investigation a “\textit{needle-in-a-haystack}” problem, with only a tiny fraction of entries related to malicious activity hidden among massive benign events.
Beyond the scarcity of attack-relevant logs, the quality of training data also impacts detection, as in \textsc{DistDet}~\cite{relatedwork_rule_graph_based:dong2023distdet}, can be contaminated by \textit{attack noise in the training period}, limiting detection accuracy.
The limited capacity of the model restricts the learning of attack patterns from log semantics tied to specific software and operating system environments, making it difficult to develop a unified model for investigation across diverse systems and software platforms. For instance, ATLAS~\cite{atlas_2021} learns attack patterns from Windows file path semantics, but entirely different path features in Linux require new feature construction and separate model training for Linux deployment. We combine the semantic features of both types of logs to train a unified model. However, as shown in ~\Cref{fig:Impact_mixed_training_data}, the average TPR decreases from 87.6\% to 57.1\%, while the average FPR decreases from 34.1\% to 17.6\%, indicating sensitivity to platform-specific features.
In addition, some methods\cite{relatedwork_rule_based:hossain2017sleuth,atlas_2021,airtag_2023,related_learning:wang2022threatrace} provide insufficient information for easily understanding attacks, since their outputs are limited to raw detection results that are hard to interpret and still mixed with benign events, forcing experts to extract scattered malicious traces buried among large volumes of benign events and to correlate and analyze them into coherent human-readable reports, which is both highly demanding and error-prone.

\section{\tool}\label{sec:approach}

\begin{figure*}[ht]
  \centering
  \includegraphics[width=1.0\textwidth]{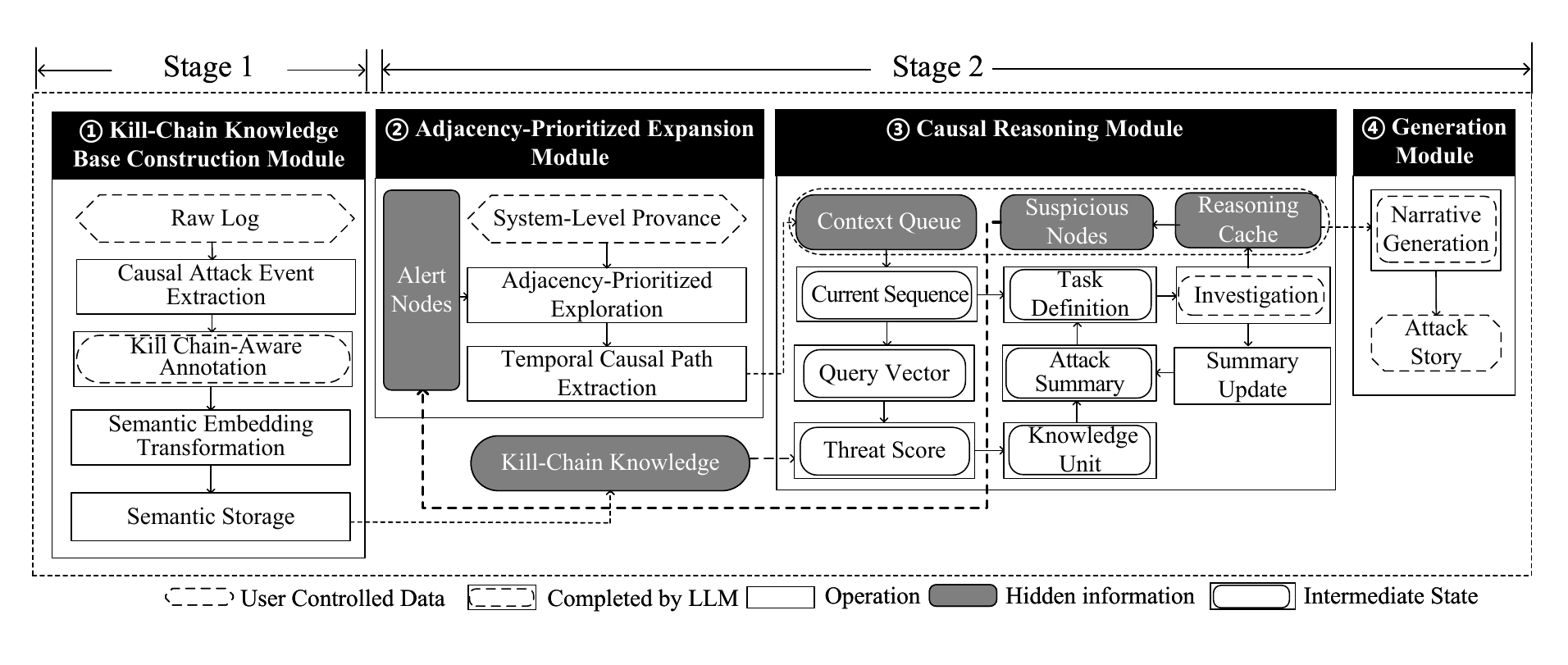}
  \caption{Overview of \tool. Stage 1 (Knowledge Base Construction) extracts and annotates causal events from known attacks to build a threat knowledge base. Stage 2 (Attack Investigation) explores the provenance graph from alert nodes, applies LLM-guided reasoning with retrieved knowledge, and generates a coherent forensic report.}
  \label{fig:overview}
\end{figure*}

\noindent As shown in \Cref{fig:overview}, \tool operates in two stages.

\noindent\textbf{Stage 1: Knowledge Base Construction.}
Raw attack logs derived from public datasets are automatically processed to extract causal events, which are annotated with Cyber Kill Chain phases (\Cref{sec:bg:killchain}) and encoded into semantic embeddings. The resulting annotated sequences are stored as structured knowledge units in a threat knowledge base (\Cref{sec:approach:knowledgebase}). This stage equips the LLM with phase-aware adversarial knowledge, avoiding labor-intensive manual annotation and ensuring that subsequent reasoning can recognize complex attack behaviors and adapt to emerging attack techniques in a timely manner without retraining.

\noindent\textbf{Stage 2: Attack Investigation.}  
Given logs from a target host, \tool constructs a provenance graph and initiates investigation from an alert event through a mutually iterative process.
The \textit{Adjacency-Prioritized Expansion Module} polls nodes from the suspicious node queue, builds their adjacency subgraphs, and converts them into time-ordered event sequences (\Cref{sec:approach:extraction}). This step keeps the investigation focused on relevant regions of the graph while avoiding dependency explosion and mitigating the constraints imposed by the limited context window.
Concurrently, the \textit{Causal Reasoning Module} retrieves knowledge units for suspicious event sequences, infers attack steps and pushes newly discovered malicious nodes to suspicious node queue for further expansion (\Cref{sec:approach:reasoning}). 
By retrieving phase-aware context from the Kill Chain knowledge base, which provides guidance for LLM reasoning, this module recognizes complex behaviors in emerging attacks and progressively reconstructs multi-phase scenarios with semantic and temporal consistency.
This feedback loop tightly couples exploration and reasoning: local suspicious contexts are extracted for analysis, while reasoning results continuously drive further exploration. 
Finally, the \textit{Generation Module} (\Cref{sec:approach:generation}) consolidates all intermediate reasoning outputs into a coherent, analyst-ready report that presents the reconstructed attack chain in chronological and causal order.

\begin{table}[tbp]
\footnotesize
\def\arraystretch{1.1}
\centering
\caption{Key Symbol Notations}
\label{tab:themis-notations}
\begin{tabular}{@{} ll @{}}
\hline
\textbf{Notation} & \textbf{Description} \\
\hline
$l \in \mathcal{L}$ & Individual log entry and system log set \\
$E_{\text{mal}}$ & Set of malicious entities \\
$G_{\text{prov}} = (V_{\text{prov}}, E_{\text{prov}})$ & Provenance graph \\
$v \in V_{\text{prov}}$, $e \in E_{\text{prov}}$ & System entity (node) and system event (edge) \\
$\mathcal{K} = \{k_0, \ldots, k_n\}$ & Phase-aware annotated sequences \\
$\mathcal{R}$ & Kill-Chain knowledge base\\
$G_{\text{adj}}(v)$ & Adjacency subgraph of $v$ induced from $G_{\text{prov}}$ \\
$\mathcal{S}_{\text{ctx}}^{v} = [s_1, s_2, \dots, s_n]$ & Local suspicious contexts of $v$ \\
$Q_{\text{sus}}$ & Suspicious node queue \\
$Q_{\text{ctx}}$ & Context queue for suspicious event sequences \\
$\mathcal{A}^{(t)} = \langle \mathcal{M}^{(t)}, \mathcal{B}^{(t)}, \mathcal{H}^{(t)} \rangle$ & Reasoning cache entry\\
\hline
\end{tabular}
\end{table}

Together, these modules form an end-to-end pipeline that integrates structured threat knowledge with iterative reasoning, enabling accurate, explainable, and analyst-friendly attack investigation. For clarity and ease of reference, we summarize the key notations frequently used in this section in Table~\ref{tab:themis-notations}.

\subsection{Kill Chain Knowledge Base Construction}
\label{sec:approach:knowledgebase}

The dynamically updatable Kill Chain knowledge base is central to \tool and is built by the Kill Chain Knowledge Base Construction Module (\Cref{fig:overview}). Given a known attack scenario, the module processes raw system logs, extracts attack-relevant events, and aligns them with the appropriate Kill Chain phases (\Cref{sec:bg:killchain}). For each phase, it generates metadata including the phase label, a behavior summary, key malicious entities, and predicted links to adjacent phases. These events and metadata are stored as structured knowledge units and indexed with retrieval indices, enabling efficient querying, providing phase-aware context while reducing the risk of hallucination.

\noindent\textbf{Malicious Events Extraction.}
To construct the knowledge base, we collect known attack scenarios, each consisting of a complete system log set $\mathcal{L}$ and a set of malicious entities $E_{\text{mal}}$. Following the standard preprocessing approach used in prior work~\cite{atlas_2021}, each log entry $l \in \mathcal{L}$ is represented as a \textit{subject–action–object} triple, as described in \Cref{sec:bg:provenance_data}. These triples contain both benign and malicious entities, such as processes, files, and IP addresses.
For each attack scenario $\langle \mathcal{L}, E_{\text{mal}} \rangle$, whether collected initially or added later through dynamic updates, \tool extracts meaningful behavior by retaining only log entries that are causally or semantically related to at least one entity in $E_{\text{mal}}$. We define a binary relevance function $\texttt{rel}(l, E_{\text{mal}})$, which returns 1 if log entry $l \in \mathcal{L}$ is associated with any entity in $E_{\text{mal}}$. The attack-relevant trace set is then defined as:
\begin{equation}
\mathcal{T} = \{ l \in \mathcal{L} \mid \texttt{rel}(l, E_{\text{mal}}) = 1 \}.
\end{equation}
This filtering process discards irrelevant system activity and preserves only the causal events required to reconstruct the attack chain. By supporting dynamic updates, the knowledge base can continuously incorporate new attack scenarios without retraining, ensuring adaptability to emerging threats.

\noindent\textbf{Phase-Aware Kill Chain Annotation.}
For each event trace $\mathcal{T}$ extracted from an attack scenario $\langle \mathcal{L}, E_{\text{mal}} \rangle$, \tool aligns the events with Kill Chain phases and generates metadata for each segment leveraging an LLM. An annotated phase unit is defined as $k_i = (m_i, t_i)$, where metadata $m_i$ is a four-tuple $\langle \textit{Phase, Behavior, Entities, Neighbors} \rangle$ and $t_i \subseteq \mathcal{T}$ is a temporally ordered event sequence. The metadata fields are defined as follows:

\smallskip
\noindent\textbullet\enspace\textit{Phase}: the Kill Chain phase (e.g., \textit{Deliver, Exploitation}) inferred through LLM reasoning guided by prior knowledge, providing phase-specific context for subsequent analysis.

\smallskip
\noindent\textbullet\enspace\textit{Behavior}: a natural-language summary of the observed activity, generated by the LLM to explain the attack behavior in human-readable form.

\smallskip
\noindent\textbullet\enspace\textit{Entities}: the key malicious entities involved, highlighted to help the LLM focus on relevant entities during reasoning.

\begin{figure}[h]
  \centering
  \includegraphics[width=0.45\textwidth,trim=1.4cm 2.5cm 0cm 0.5cm, clip]{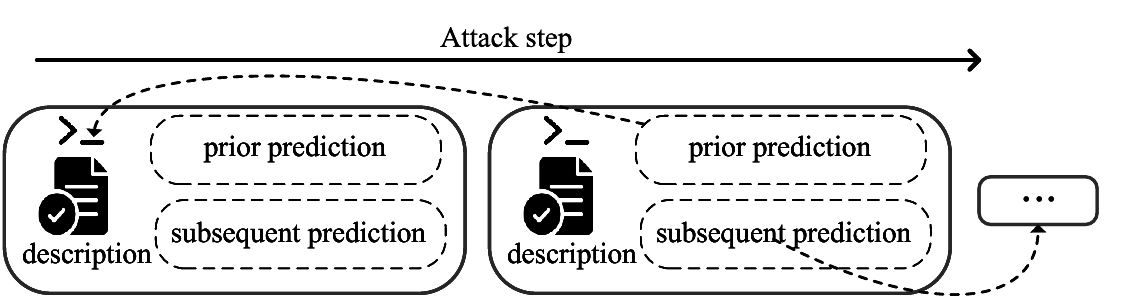}
  \caption{Example of phase-aware annotation. Each sequence includes a summary of the preceding step and a prediction of the next phase, ensuring continuity across Kill Chain phases and supporting investigations that may begin mid-attack.}
  \label{fig:guide_summary}
\end{figure}

\smallskip
\noindent\textbullet\enspace\textit{Neighbors}: predicted descriptions of both preceding and subsequent phases, ensuring that each unit is contextually grounded and interpretable even when investigation begins mid-attack (illustrated in \Cref{fig:guide_summary}).

Segmentation and annotation are performed using an LLM guided by a structured prompt $P_{\text{kill}}$ (\Cref{llm:prompt:annotate}), designed to (i) align events with appropriate Kill Chain phases, (ii) identify both malicious and benign entities and describe their behaviors, and (iii) generate concise behavioral summaries with links to adjacent phases to preserve continuity. Given $\mathcal{T}$, $E_{\text{mal}}$, and $P_{\text{kill}}$, the LLM produces a collection of phase-specific annotated sequences:
\begin{equation}
\mathcal{K} = \{ k_{0}, \dots, k_{n} \} = \texttt{LLM}\left(\mathcal{T}, E_{\text{mal}}, P_{\text{kill}} \right).
\end{equation}
This phase-aware segmentation is preferred over generic, context-free chunking methods because the Kill Chain provides a structured framework that reflects the logical progression of real-world attacks and enhances LLM reasoning.

\noindent\textbf{Semantic Embedding and Index Construction.}
Given a collection of phase-aware annotated sequences $\mathcal{K} = \{ k_{0}, \dots, k_{n} \}$, each $k_i = (m_i, t_i)$ is processed by embedding its temporally ordered event sequence $t_i$ into dense vectors. These vectors serve as retrieval indices that enable efficient similarity-based search over both the attack behavior sequences and their metadata $m_i$. Logs from different platforms exhibit semantic differences, yet the embedding space maintains clear separation across platforms and coherent organization within each platform, ensuring effective retrieval and reasoning.

To construct retrieval indices, each event sequence $t_i$ is divided into smaller subsequences when necessary, due to the token limit of the embedding model $\mathcal{E}(\cdot)$. Specifically, $t_i$ is split into units $t_i = \{ u_1, u_2, \dots, u_j \}$, where each $u_j$ contains at most $N$ events ($N=20$ in our implementation, balancing semantic completeness with model capacity). Each subsequence $u_j$ inherits the metadata $m_i$ of its parent sequence. The embedding model then produces dense vector representations:
\begin{equation}
\mathcal{R}^{(i)} = \{ r_j \mid u_j \in t_{i}, r_j = \mathcal{E}(u_j) \}.
\end{equation}
Each embedding $r_j$ is stored in the knowledge base $\mathcal{R}$ together with its original subsequence $u_j$ and metadata $m_i = \langle \textit{Phase, Behavior, Entities, Neighbors} \rangle$. This structure~enables similarity-based retrieval while preserving phase-aware semantic and contextual annotation of each knowledge unit.

\subsection{Suspicious Context Exploration}
\label{sec:approach:extraction}

\begin{figure}[htb]
\centering
\includegraphics[width=0.48\textwidth,trim=0cm 1.2cm 0cm 0cm, clip]{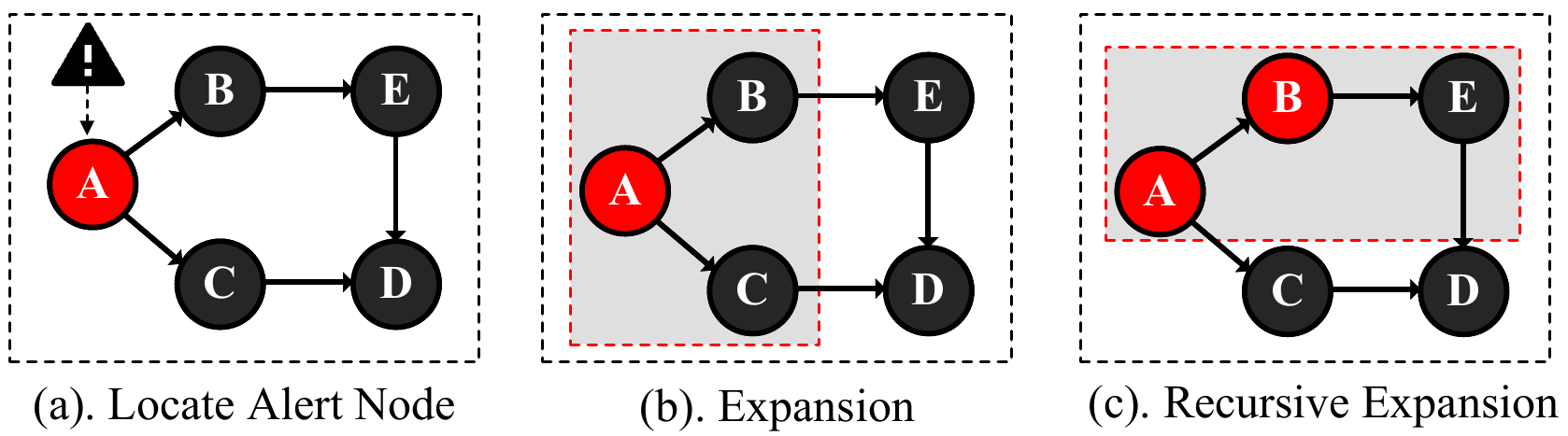}
\begin{subfigure}{0.14\textwidth}
    \caption{Initial alert}
    \label{fig:graph_expand:init}
\end{subfigure}%
\begin{subfigure}{0.18\textwidth}
    \caption{Exploring neighbors}
    \label{fig:graph_expand:exp}
\end{subfigure}%
\begin{subfigure}{0.16\textwidth}
    \caption{Graph expansion}
    \label{fig:graph_expand:expand}
\end{subfigure}%
\caption{Illustration of iterative causal reasoning. (a) Identify initial alert node \texttt{A}. (b) Extract local suspicious context by exploring adjacent nodes \texttt{B} and \texttt{C}. (c) Expand the investigation when node \texttt{B} is identified as suspicious.}

\label{fig:graph_expand}
\end{figure}

With the Kill Chain knowledge base $\mathcal{R}$ in place, \tool initiates attack investigation by transforming the collected logs (\Cref{threat_model}) into a provenance graph and using the initial alert event as the entry point (\Cref{fig:graph_expand:init}). 
The Adjacency-Prioritized Expansion Module then iteratively polls a node from the \textit{suspicious node queue}, extracts its adjacency subgraph (\Cref{fig:graph_expand:exp}), converts the subgraph into temporally ordered event sequences, and places them into the \textit{context queue}. 
By localizing analysis around suspicious nodes and converting graph fragments into bounded sequences, this exploration mechanism ensures that LLM-based reasoning remains both focused and scalable, effectively mitigating context-length limitations while preserving relevant causal evidence.

\noindent\textbf{Provenance Graph Assembly.} 
Before exploration begins, the module constructs a provenance graph from the logs of the target host under investigation (\Cref{sec:bg:provenance_data}).
This graph may span several days of activity and typically contains hundreds of thousands of events across tens of thousands of nodes.
As described in \Cref{sec:bg:investigation} and \Cref{sec:approach:knowledgebase}, the collected logs are parsed into structured \textit{subject–action–object} triples, which are assembled into a provenance graph 
$G_{\text{prov}} = (V_{\text{prov}}, E_{\text{prov}})$.
Each node $v \in V_{\text{prov}}$ represents a system entity, and each edge $e \in E_{\text{prov}}$ denotes an interaction between entities with a timestamp.
The resulting graph captures fine-grained system activities over time, providing a compact and semantically rich representation of host behavior.

\noindent\textbf{Local Suspicious Context Extraction.}
To ensure that \tool remains focused on the most relevant causal evidence, the Adjacency-Prioritized Expansion Module iteratively extracts the local context of suspicious nodes for subsequent reasoning. 
The investigation begins with a security alert event, whose corresponding alert node $v_{\text{alert}} \in V_{\text{prov}}$ is identified (\Cref{fig:graph_expand:init}) and placed into the suspicious node queue $Q_{\text{sus}}$.

At each iteration, the module polls a suspicious node $v$ from $Q_{\text{sus}}$ and constructs its one-hop neighborhood:
\begin{equation}
V_{\text{adj}}(v) = \{ v \} \cup \{ u\!\in\!V_{\text{prov}}\!\mid\!(v, u)\!\in\!E_{\text{prov}} \vee (u, v)\!\in\!E_{\text{prov}} \}.
\label{eq:neighbors}
\end{equation}
This set includes $v$ and all nodes directly connected to it, thereby capturing its immediate causal context. The adjacency subgraph of node $v$ is then defined as:
\begin{equation}
G_{\text{adj}}(v) = G_{\text{prov}}[V_{\text{adj}}(v)],
\end{equation}
which restricts analysis to the one-hop neighborhood of $v$ and avoids unnecessary traversal into unrelated regions (\Cref{fig:graph_expand:exp}), preserving the local causal context most relevant to the current suspicious node.

Since LLMs operate more effectively on sequential data than on graph structures, $G_{\text{adj}}(v)$ is converted into a temporally ordered sequence of events. The edge set of this subgraph~is
\begin{equation}
E_{\text{adj}}(v) = \{ (u,w) \in E_{\text{prov}} \mid u,w \in V_{\text{adj}}(v) \},
\end{equation}
where each edge represents a system event. Ordering these events by their timestamps yields a sequential representation of host behavior.
To remain consistent with the procedure described in \Cref{sec:approach:knowledgebase}, the event sequence is partitioned into subsequences of bounded length $N$:
\begin{equation}
\mathcal{S}_{\text{ctx}}^{v} = [s_1, s_2, \dots, s_n], \quad s_i \subseteq E_{\text{adj}}(v) \wedge  |s_i| \leq N.
\end{equation}
Each subsequence $s_i$ is aligned with the knowledge base representation, enabling direct retrieval of phase-aware context. Finally, every $s_i \in \mathcal{S}_{\text{ctx}}^{v}$ is pushed into the context queue $Q_{\text{ctx}}$, which serves as a sliding buffer of suspicious event sequences awaiting investigation.

This design allows \tool to focus on the causal context most relevant to each suspicious node while filtering out benign background activity, thus producing semantically coherent suspicious context for subsequent causal reasoning.

\subsection{Iterative Causal Reasoning}
\label{sec:approach:reasoning}

The Causal Reasoning Module leverages the dynamic Kill Chain knowledge base $\mathcal{R}$ to retrieve phase-aware context, enabling recognition of complex behaviors in emerging attacks and the progressive reconstruction of multi-phase intrusions.
In each iteration $t$, the module pops a suspicious event sequence $s^{(t)}$ from the context queue $Q_{\text{ctx}}$, retrieves a phase-aware knowledge unit $r^{(t)}$ from the Kill Chain base, and infers the next attack progression while updating the \textit{reasoning cache} to maintain continuity across iterations. 
Newly discovered malicious nodes are pushed back to the suspicious node queue $Q_{\text{sus}}$ for further exploration (\Cref{fig:graph_expand:expand}). 
Through this feedback loop, \tool performs \textit{iterative causal reasoning} in an emerging attack scenario, progressively linking local dependencies and long-range attack progressions to produce consistent evidence for analyst-ready reports.

\noindent\textbf{Threat Knowledge Retrieval.}
To guide reasoning, the suspicious event sequence $s^{(t)}$ is first encoded into a semantic vector $q^{(t)} = \mathcal{E}(s^{(t)})$ using the pretrained embedding model introduced in \Cref{sec:approach:knowledgebase}. As described in \Cref{sec:bg:rag}, cosine similarity scores are then computed between $q^{(t)}$ and the vector representations $\mathcal{E}(r_i)$ of all knowledge units $r_i$ in the Kill Chain knowledge base $\mathcal{R}$:
\begin{equation}
\mathrm{ThreatScore}(q^{(t)}, r_i) = \frac{q^{(t)} \cdot \mathcal{E}(r_i)}{\|q^{(t)}\| \, \|\mathcal{E}(r_i)\|}.
\end{equation}
The most semantically relevant knowledge unit is retrieved as
\begin{equation}
r^{(t)} = \arg\max_{r_i \in \mathcal{R}}\;\mathrm{ThreatScore}(q^{(t)}, r_i).
\end{equation}
The retrieved unit $r^{(t)}$ includes a phase-aware annotation $\langle \textit{Phase, Behavior, Entities, Neighbors} \rangle$ together with a representative attack sequence. This contextual information serves as external domain knowledge that guides the LLM toward accurate attack-path reasoning, reducing hallucinations and ensuring phase consistency.

\noindent\textbf{Reasoning and Cache Update.}
To maintain causal continuity across iterations, \tool employs a \textit{reasoning cache} that acts as explicit memory. Each entry at iteration $t$ is stored as
\begin{equation}
\mathcal{A}^{(t)} = \langle \mathcal{M}^{(t)}, \mathcal{B}^{(t)}, \mathcal{H}^{(t)} \rangle,
\end{equation}
where $\mathcal{M}^{(t)}$ denotes the set of newly identified malicious nodes, $\mathcal{B}^{(t)}$ the inferred descriptions of attack behaviors, and $\mathcal{H}^{(t)}$ an updated summary of the reconstructed attack path.

To maintain causal continuity across iterations, the module designs a structured reasoning template $P_{\text{reasoning}}$ (\Cref{llm:prompt:reasoning}) that (i) integrates the current sequence, retrieved knowledge, and historical summary,  
(ii) distinguishes benign from malicious nodes with behavior analysis, and  
(iii) summarizes the inferred step, key entities, and evidence.
The LLM is prompted with the current event sequence $s^{(t)}$, the retrieved threat knowledge $r^{(t)}$, the historical summary $\mathcal{H}^{(t-1)}$ from the last iteration, and a structured reasoning template $P_{\text{reasoning}}$ : 
\begin{equation}
\mathcal{A}^{(t)} = \texttt{LLM}\bigl(s^{(t)}, r^{(t)}, \mathcal{H}^{(t-1)}, P_{\text{reasoning}}\bigr).
\label{eq:reasoning_input}
\end{equation}
The generated cache entry $\mathcal{A}^{(t)}$ is appended to the reasoning cache, ensuring persistence of intermediate results. Meanwhile, the updated summary $\mathcal{H}^{(t)}$ provides short-term continuity linking current reasoning step to subsequent iterations.

\noindent\textbf{Feedback to Exploration.}
To drive the next round of exploration and maintain causal continuity, the set of suspicious nodes $\mathcal{M}^{(t)}$ discovered in iteration $t$ is added back into the suspicious node queue $Q_{\text{sus}}$ and passed to the Adjacency Prioritized Expansion Module for further suspicious context exploration (\Cref{fig:graph_expand:expand}). This feedback loop tightly couples reasoning with exploration: each round of reasoning guides the next stage of context extraction, allowing \tool to iteratively refine causal reasoning and progressively reconstruct the complete attack story.


\subsection{Attack Story Generation}
\label{sec:approach:generation}

To bridge automated reasoning with practical forensic output, the Generation Module transforms the per-iteration outputs stored in the reasoning cache into an analyst-ready investigation report, ensuring that the reconstructed attack chain is both logically consistent and easy for analysts to interpret.

Once the context queue $Q_{\text{ctx}}$ has been fully processed, the module collects all reasoning outputs $[\mathcal{A}^{(1)}, \mathcal{A}^{(2)}, \ldots, \mathcal{A}^{(T)}]$ together with the latest summary $\mathcal{H}^{(T)}$ from the reasoning cache. 
To guarantee global consistency, the module constructs a narrative prompt $P_{\text{gen}}$ that (i) enforces temporal ordering of events, (ii) explicitly links causality across reasoning steps, and (iii) requires clear, analyst-readable phrasing enriched with Kill Chain annotations. 

Given these structured inputs, the final synthesis is produced by the LLM:
\begin{equation}
\mathcal{G} = \texttt{LLM}\bigl([ \mathcal{A}^{(1)}, \mathcal{A}^{(2)}, \ldots, \mathcal{A}^{(T)} ],\, \mathcal{H}^{(T)},\, P_{\text{gen}}\bigr),
\label{eq:gen_final}
\end{equation}
where $\mathcal{G}$ denotes the global report that chronologically and causally connects the intrusion steps, augmented with natural-language explanations and phase labels. Here, $T$ represents the total number of reasoning iterations, $P_{\text{gen}}$ is a structured generating template (\Cref{llm:prompt:generation})

By consolidating the reasoning results into a single global narrative, we deliver a clear, comprehensive, and actionable attack report suitable for real-world incident response.

\section{Evaluation}\label{sec:Evaluation}

\begin{table*}[htbp]
\setlength{\tabcolsep}{0.4em}
\centering
\caption{Overview of Datasets with Attack Types and Alert Clues}
\label{tab:dataset_detail}
\resizebox{\linewidth}{!}{
\begin{tabular}{llc l c r r r r r}
\toprule
\textbf{ID} & \textbf{Attack Type} & \textbf{Lateral Movement} & \textbf{Alert Clues Type} & \textbf{\#Comp. Proc. } & \textbf{Target OS} & \textbf{Events} & \textbf{Nodes} & \textbf{\#\%Malicious} & \textbf{Log Size (MB)} \\
\midrule
S1 & Web compromise & \ding{55} & Domain, IP, Executable& \ding{55}& Windows & 95.0K & 7,468 & 6.46\% & 382  \\
S2 & Malvertising & \ding{55} & Domain, IP, Executable&  \ding{55} & Windows & 397.9K & 34,021 & 4.30\% & 1015  \\
S3 & Spam campaign & \ding{55} & Domain, IP, Executable, Doc &  \ding{55}& Windows & 128.3K & 8,998 & 12.12\% & 522  \\
S4 & Pony campaign & \ding{55} & Domain, IP, Executable, Doc &  \ding{55}& Windows & 125.6K & 13,037 & 16.17\% & 499  \\
M1 & Web compromise & \ding{51} & Domain, IP, Executable &  \ding{55}& Windows & 251.6K & 17,599 & 4.06\% & 711,102  \\
M2 & Phishing & \ding{51} & Domain, IP, Executable &  \ding{55}& Windows & 284.3K & 24,496 & 13.68\% & 671,112  \\
M3 & Malvertising & \ding{51} & Domain, IP, Executable &  \ding{55}& Windows & 334.1K & 24,481 & 11.60\% & 336,138  \\
M4 & Monero miner & \ding{51} & Domain, IP, Executable, Doc &  \ding{55}& Windows & 258.7K & 15,409 & 3.80\% & 533,91  \\
M5 & Pony campaign & \ding{51} & Domain, IP, Executable, Doc&  \ding{55} & Windows & 586.6K & 35,709 & 5.33\% & 726,113  \\
M6 & Spam campaign & \ding{51} & Domain, IP, Executable, Doc &  \ding{55}& Windows & 354.0K & 19,666 & 4.18\% & 55,142  \\
DL & Data Exfiltration & \ding{55} & IP, File, Executable&  \ding{55}& Linux & 506.1K & 20,288 & 0.07\% & 112  \\
VF & Malware & \ding{55} & IP, File, Executable&  \ding{55}& Linux & 221.1K & 6,908 & 0.001\% & 68  \\
SP & Bashdoor & \ding{55} & IP, File, Executable & \ding{55}& Linux & 134.6K & 3,659 & 0.003\% & 48 \\
EB-P1 & Gamaredon & \ding{55} & IP, Executable, DLL & \ding{51}& Windows & 218K & 17,421 & 0.499\% & 621 \\
EB-P2 & Gamaredon & \ding{55} & IP, Executable, Script & \ding{51} & Windows & 442.7K & 22,066 &0.501\% & 344  \\
\bottomrule
\end{tabular}
}
\caption*{\footnotesize DL, VF, and SP denote the DataLeak, VPNFilter, and Shellshock-Penetration datasets. S1–S4 and M1–M6 correspond to single-host and multi-host attack scenarios, respectively, while EB-P1 and EB-P2 are extended benchmark cases based on Gamaredon attacks.
\textbf{Comp. Proc.} denotes \emph{Compromised System Process}.
}
\end{table*}

Guided by the threat model in \Cref{threat_model}, we evaluate \tool across five dimensions. We first assess its effectiveness in reconstructing complete attack scenariosand identifying malicious behaviors (\Cref{effectiveness}). We then examine its generalization capability, demonstrating strong generalization across heterogeneous logs from different operating systems without retraining or feature engineering, as well as adaptability to diverse alert types such as domains, processes, and IP addresses (\Cref{Generalization Capability}). We further evaluate its efficiency in processing attack scenarios within practical time and token budgets (\Cref{sec:efficiency}), followed by an analysis of the impact of design choices (\Cref{sec:ablation}). Finally, we present a case study that illustrates the application of \tool in a realistic investigation setting~(\Cref{sec:case study}).

\subsection{Evaluation Setup}

\noindent\textbf{LLMs.} \tool uses \texttt{DeepSeek-R1}, an open-source model trained with reinforcement learning~\cite{r1:guo2025deepseek}. 
This model supports step-by-step reasoning, which enhances causal analysis of attacks. 
We additionally evaluate with \texttt{GPT-4o} and \texttt{GPT-o1} models; unless otherwise specified, \texttt{DeepSeek-R1} is used by default. 
All experiments invoke the models via their APIs.

\smallskip
\noindent\textbf{RAG System.} \tool adopts a retrieval-augmented generation design consisting of a retriever and a vector database. The retriever embeds event sequences using \texttt{text-embedding-v1} and computes similarity with cosine, Euclidean, or inner product distance. Retrieved embeddings and their metadata are stored in Milvus~\cite{milvusMilvusHighPerformance}, a high-performance vector database.

\smallskip
\noindent\textbf{Datasets.} We evaluate \tool on datasets from both Windows and Linux platforms, each annotated with ground-truth malicious entities (\Cref{tab:dataset_detail}).

\noindent\textbullet\enspace\textit{ATLAS Dataset}: Ten APT scenarios derived from real-world reports, including single-host cases (S1-S4) and multi-host cases with lateral movement (M1-M6). Each averages 249K events interleaved with benign activities (e.g., browsing, video playback), with logs labeled per host (e.g., M1h1).

\noindent\textbullet\enspace\textit{EB Attacks Dataset}: Two single-host scenarios (EB-P1 and EB-P2) based on real-world APT reports~\cite{eteralBlue_broadcomEndpointProtection,wannacrty_wikipediaWannaCryRansomware}, executed on Windows 7 Virtual Machines with distinct environments. Both exploit CVE-2017-6334 via shellcode injection, followed by persistence through service registration (EB-P1) or scheduled task execution (EB-P2). Each lasts about one hour, followed by 24 hours of benign activity for realism.

\noindent\textbullet\enspace\textit{DepImpact Datasets}: Three Linux scenarios, \texttt{Dataleak}, \texttt{Vpnfilter}, and \texttt{Shellshock}, built from audit logs of realistic attack simulations following \textsc{DepImpact} setup~\cite{relatedwork_rule_based:fang2022back}.

\smallskip
\noindent\textbf{Evaluation Protocol.} We consider \atlas~\cite{atlas_2021} and \airtag~\cite{airtag_2023} as representative baselines. We noticed that the test set of \airtag contains label-related words in the logs that explicitly indicate whether a sample is positive or negative, which would not occur in real-world environments. Given this sensitivity, we exclude \airtag from our main comparisons and focus on \atlas, following its experimental setup and report the average performance across different start points. We have contacted the \airtag authors for clarification but have not yet received a response.

\smallskip
\noindent
\textbf{Evaluation Metrics.}
We adopt standard metrics based on True Positive (TP), False Positive (FP), True Negative (TN), and False Negative (FN).

\noindent\textbullet\enspace\textit{TPR (True Positive Rate)}: $ \text{TPR} = \tfrac{TP}{TP+FN} $, proportion of malicious events correctly detected.

\noindent\textbullet\enspace\textit{FPR (False Positive Rate)}: $ \text{FPR} = \tfrac{FP}{FP+TN} $, proportion of benign events misclassified as malicious.

\noindent\textbullet\enspace\textit{Balanced Accuracy}: evaluates performance under class imbalance, where benign events vastly outnumber malicious ones. It is defined as $ \text{Balanced Accuracy} = \tfrac{\text{TPR} + (1 - \text{FPR})}{2} $.

\noindent\textbullet\enspace\textit{Token Cost}: total tokens consumed during the entire investigation process.

\smallskip
\noindent\textbf{Platform.} Experiments are conducted on a server running 64-bit Ubuntu 20.04 LTS with an Intel(R) Xeon(R) Silver 4214 CPU @ 2.20GHz, 692GB RAM, and four Tesla V100 GPUs (32GB each).

\begin{table*}[t]
\setlength{\tabcolsep}{1.1em}
\centering
\caption{Results of Using Different Start Points in ATLAS and \tool.}

\label{tab:comparison}
\resizebox{\linewidth}{!}{
\begin{tabular}{c|cc cc cc|cc cc cc} 
\toprule
\multirow{3}{*}{Dataset} & \multicolumn{6}{c|}{\atlas} & \multicolumn{6}{c}{\tool}\\\cline{2-13}
& \multicolumn{2}{c}{Domain Name} & \multicolumn{2}{c}{IP Address} & \multicolumn{2}{c|}{Payload} & \multicolumn{2}{c}{Domain Name} & \multicolumn{2}{c}{IP Address} & \multicolumn{2}{c}{Payload}  \\
                         & TPR   & FPR                   & TPR   & FPR                     & TPR    & FPR                 & TPR    & FPR                  & TPR    & FPR                    & TPR    & FPR                  \\\cline{1-13}
S1  & 99.7\% & 0.0\%                 & 6.1\%  & 0.0\%                   & 100.0\% & 47.2\%             & 100.0\% & 0.0\%                & 100.0\% & 0.0\%                  & 100.0\% & 0.0\%                \\
S2  & 100.0\% & 0.0\%                 & 100.0\% & 81.2\%                 & 100.0\% & 92.8\%             & 100.0\% & 0.0\%                & 100.0\% & 0.0\%                  & 100.0\% & 0.0\%                \\
S3  & 78.5\% & 0.9\%                 & 99.1\% & 0.0\%                   & 100.0\% & 84.5\%             & 100.0\% & 0.0\%                & 100.0\% & 0.0\%                  & 100.0\% & 0.0\%                \\
S4  & 99.7\% & 0.0\%                 & 100.0\% & 44.7\%                 & 100.0\% & 81.2\%             & 100.0\% & 0.0\%                & 100.0\% & 2.5\%                  & 100.0\% & 0.0\%                
         \\ 
\midrule
Average & 94.5\% & 0.2\%                 & 76.3\% & 31.5\%                  & 100.0\% & 76.4\%             & 100.0\% & 0.0\%                & 100.0\% & 0.6\%                  & 100.0\% & 0.0\%                      \\
\bottomrule
\end{tabular}
}

\end{table*}

\subsection{Effectiveness of Attack Reconstruction}
\label{effectiveness}
We evaluate the effectiveness of \tool in comparison with \atlas under two settings: (i) \textit{Similar Exploitation Attacks}, represented by the ATLAS dataset, where operating system, applications, and user behaviors are consistent across scenarios; and (ii) \textit{Novel Shellcode Injection Attacks}, represented by EB-P1 and EB-P2, where diverse environments and unseen exploitation techniques are introduced.

\begin{figure}[h]
    \centering
    \includegraphics[width=0.85\linewidth]{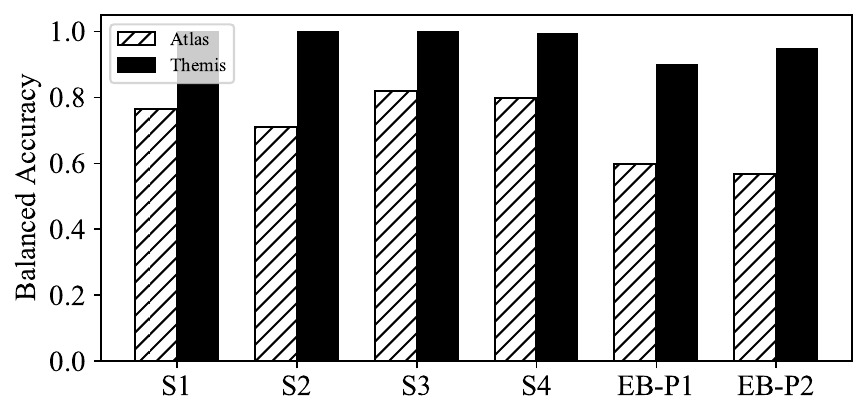}
    \caption{Balanced Accuracy comparison between \atlas and \tool across single-host scenarios.}
    \label{fig:single_dataset.png}
\end{figure}

\begin{figure}
    \centering
    \includegraphics[width=0.95\linewidth]{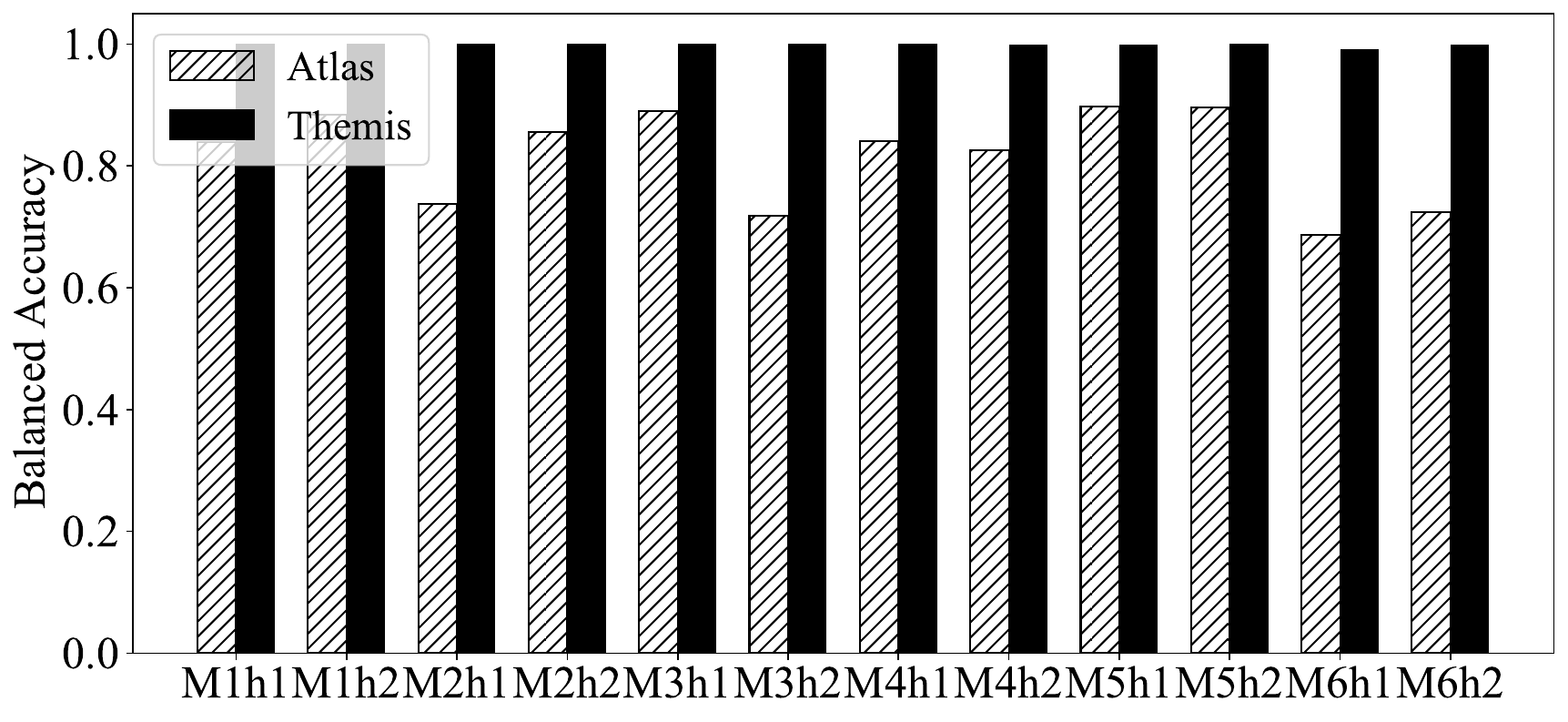}
    \caption{Balanced Accuracy comparison of \atlas and \tool across multi-host attack scenarios (M1--M6).}
    \label{fig:Atlas_multi}
\end{figure}

\noindent
\textbf{Similar Exploitation Attack.}
In single-host scenarios S1–S4, all attack activities are confined to one machine, providing a controlled baseline for comparison. As shown in \Cref{fig:single_dataset.png}, \atlas achieves Balanced Accuracy between 71.0\% and 82.0\%, indicating limited ability to fully reconstruct attack traces. By contrast, \tool consistently delivers highly accurate results, with Balanced Accuracy of 100.0\% (S1), 99.9\% (S2), 99.9\% (S3), and 99.5\% (S4). 
We then consider multi-host scenarios (M1–M6), which involve lateral movement. As shown in \Cref{fig:Atlas_multi}, \atlas achieves Balanced Accuracy between 68.6\% and 89.7\%, with significant variation across cases. In sharp contrast, \tool consistently exceeds 99.9\% across all hosts. For example, in M6h1 \atlas achieves 68.7\%, while \tool reaches 99.1\% . Even in scenarios where \atlas performs relatively well (e.g., M5h1: 89.7\%), \tool improves further to 99.9\%. These results demonstrate that \tool not only compensates for \atlas's weaknesses in difficult cases but also strengthens effectiveness in favorable conditions, showing resilience against complex multi-host attack behaviors.

\noindent
\textbf{Novel Shellcode Injection Attack.}
We evaluate adaptability on EB-P1 and EB-P2, which introduce diverse configurations, user behaviors, and stealthy shellcode injection. Both \atlas (trained on S1-S4) and \tool (knowledge base built from the same scenarios) are directly applied without retraining or updates,and results are averaged across multiple alert entry points. As shown in \Cref{fig:single_dataset.png}, \atlas achieves Balanced Accuracy of 59.8\% (EB-P1) and 56.7\% (EB-P2), underscoring its difficulty in handling previously unseen attack patterns. By contrast, \tool attains 90.3\% and 94.8\% respectively, corresponding to absolute gains of 30.4\% and 38.1\%. These results highlight that \tool effectively adapts to novel exploitation techniques without retraining or knowledge-base updates, maintaining high investigation accuracy across diverse and evolving attack settings.

\subsection{Generalization Capability}
\label{Generalization Capability}
We evaluate the generalization capability of \tool across different platforms and diverse alert types. For platform generalization, we test on Linux datasets, which differ in semantics and log formats from Windows. For alert generalization, we initiate investigations from different types of alerts such as processes, domains, and IP addresses. These experiments demonstrate that \tool reliably reconstructs attack scenarios across heterogeneous platforms and diverse alert types.

\begin{figure}
    \centering
    \includegraphics[width=0.85\linewidth]{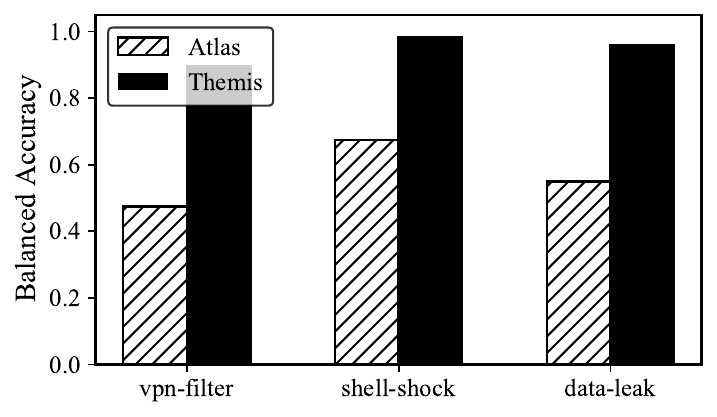}
    \caption{Balanced Accuracy comparison of \atlas and \tool across three Linux attack scenarios.}
    \label{fig:linux_dataset.png}
\end{figure}

\noindent\textbf{Generalization to Platform.}
On Linux datasets (\texttt{Dataleak}, \texttt{VPNFilter}, and \texttt{Shellshock}), we evaluate \atlas using the public \airtag implementation~\cite{airtag_2023}. \atlas builds a vocabulary from training logs and tokenizes system events into discrete identifiers, which introduces platform-specific dependencies (e.g., Linux events such as \texttt{execve}, \texttt{write}, and \texttt{recvfrom} differ from Windows semantics). \atlas achieves Balanced Accuracies of 47.5\%, 67.5\%, and 55.0\% on the three datasets, highlighting limited adaptability to heterogeneous environments. In contrast, \tool achieves 89.5\%, 98.3\%, and 95.8\%, yielding absolute improvements of 42.0\%, 30.8\%, and 40.9\%. These consistent gains confirm that \tool maintains high-quality reconstruction across platforms without retraining or manual feature engineering, demonstrating generality against semantic and structural diversity in log data.

\noindent\textbf{Generalization to Alert Types.}
We further test generalization by varying the types of alert-related entities used as investigation entry points (Table~\ref{tab:comparison}).

\noindent
\textit{Domain Name.} Both methods achieve similar performance, as malicious domains are consistently resolved to IP addresses and exhibit limited benign interactions, making them straightforward to distinguish.

\noindent
\textit{Payload.} While both achieve high TPR, \atlas suffers from high FPR due to frequent benign interactions (e.g., with \texttt{svchost.exe}, \texttt{dllhost.exe}). \tool leverages contextual reasoning to disambiguate them, substantially reducing FPR.

\noindent
\textit{IP Address.} \atlas shows unstable performance, with a sharp TPR drop in S1 due to the excessive number of IP-related events and consistently elevated FPR (e.g., 81.2\% in S2). In contrast, \tool maintains high TPR and low FPR across all scenarios, demonstrating generality with varied entry points.

\subsection{Efficiency}
\label{sec:efficiency}

We evaluate the efficiency of \tool across four representative workloads: single-host scenarios (S1–S4), EternalBlue-based intrusions (EB), cross-host attack chains (Multi), and DepImpact on Linux logs (Dep). This evaluation considers both processing time and token usage, providing insight into how workload characteristics affect resource consumption.

\begin{table}[h]
\centering
\caption{Cost Performance Metrics}
\label{tab:cost_metrics}
\setlength{\tabcolsep}{3pt}  
\resizebox{\linewidth}{!}{
\begin{tabular}{lccccc}
\toprule
\textbf{Workload} & \textbf{Time (h)} & \textbf{Answer (k)} & \textbf{Prompt (k)} & \textbf{Reasoning (k)} & \textbf{Total (k)} \\
\midrule
Single   & 1.0 & 25.5 & 77.7 & 41.9 & 145.0 \\
EB  & 3.5 & 142.3 & 200.4 & 97.8 & 440.5 \\
Multi   & 1.9 & 78.9  & 109.7 & 52.9 & 241.5 \\
Dep & 1.3 & 52.8  & 74.4  & 36.3 & 163.5 \\
\midrule
Avg & 1.9 & 74.9 & 115.6 & 57.2 & 247.6 \\
\bottomrule
\end{tabular}
}
\end{table}

\smallskip
\noindent\textbf{Processing Time.}
On average, investigations complete within 1.9 hours. Single-host cases are the fastest at 1.0 hour, while EB incurs the longest runtime of 3.5 hours due to extended attack duration and higher log volume. Multi-host and Dep workloads require 1.9 and 1.3 hours, respectively, reflecting the tradeoff between cross-host traversal and platform-specific log complexity. Although not negligible, these runtimes are acceptable for offline forensic analysis, where accuracy and completeness outweigh real-time requirements.

\smallskip
\noindent\textbf{Token Usage.}
The average token consumption is 247.6k per case, comprising 115.6k for prompts, 57.2k for reasoning, and 74.9k for final report generation. EB is the most expensive at 440.5k tokens, driven by persistent adversarial techniques and frequent relocation of malicious executables, which extend reasoning paths and inflate costs. In contrast, single-host cases require only 145.0k tokens. Multi-host and Dep workloads consume 241.5k and 163.5k tokens, respectively. These results show that workloads with dense attack graphs or cross-host propagation significantly increase reasoning overhead.

\subsection{Impact of Design Choices}
\label{sec:ablation}

To understand how different design choices influence the effectiveness of \tool, we conduct an ablation study on the single-host scenarios (S1–S4). We begin with baseline designs that progressively add key components (i.e., graph exploration strategies and knowledge augmentation), and then examine sensitivity to different LLMs, additional knowledge units, and similarity metrics. This evaluation quantifies the contribution of each technique and parameter to overall performance.

\begin{table}[ht]
\centering
\caption{Results of Different Design Choices}
\label{tab:different_method}
\setlength{\tabcolsep}{0.3em}
\begin{tabular}{lrrrr}
\toprule
\textbf{Method} & \textbf{TPR} & \textbf{FPR}  & \textbf{Time (h)} & \textbf{Token (k)} \\
\midrule
Graph Traversal   & 100.0\% & 71.3\% & 21.1  & 3,371 \\
Adjacency-Graph   & 99.8\%  & 36.2\% & 12.2  & 1,112 \\
Unstructured Base & 98.9\%  & 6.3\%  & 1.7   & 235   \\
\midrule
\tool          & 100\% & 0.2\% &1.0 &145\\
\bottomrule
\end{tabular}
\end{table}

\smallskip
\noindent\textbf{Graph Traversal.}
We first consider exhaustive graph traversal, which traces all forward and backward dependencies from an alert node and directly feeds the resulting sequences to the LLM, similar to \atlas~\cite{atlas_2021}. As shown in Table~\ref{tab:different_method}, this method achieves a TPR of 100\% but suffers from a high FPR (71.3\%) due to the inclusion of many benign events. It is also highly inefficient, requiring 21.1 hours and 3,371k tokens. Since its prior knowledge often associates such processes with auxiliary roles in attacks, the LLM frequently misclassifies benign processes such as \texttt{svchost.exe} and \texttt{dllhost.exe}, further inflating false positives and cost.

\smallskip
\noindent\textbf{Adjacency-Prioritized Expansion.}
This design restricts graph exploration to local neighborhoods without using contextual knowledge. Event sequences from structurally adjacent nodes are directly passed to the LLM for reasoning.

As shown in Table~\ref{tab:different_method}, this approach maintains a high TPR (99.8\%) while reducing FPR to 36.2\%, with runtime shortened to 12.2 hours and token usage lowered to 1,112k, demonstrating substantial improvements over exhaustive traversal. However, the absence of knowledge guidance leads the LLM to over-predict malicious behavior. For example, in attacks delivered via Word documents, \texttt{winword.exe} is often misclassified as malicious, triggering unnecessary dependency tracing to benign processes such as \texttt{services.exe}, which reintroduces dependency explosion and irrelevant events.

\smallskip
\noindent\textbf{Unstructured Knowledge Augmentation.}
We evaluate adjacency-prioritized expansion combined with a knowledge base that stores attack behaviors using simple annotations rather than Kill-Chain organization. Compared with using no knowledge base, this design significantly improves precision: TPR remains high at 98.9\%, while runtime and token usage drop to 1.7 hours and 235k, respectively. However, the lack of phase-aware organization sometimes causes temporal misalignment, leading the LLM to miss document-based exploits or misclassify benign components such as \texttt{plugin-container.exe}. Compared with the Kill Chain–organized knowledge base, this setting still yields higher error rates, with FPR reduced only to 6.3\%.



\noindent
\textbf{Impact of Different LLMs.}
We next evaluate the effect of using different LLM backends. As shown in \Cref{tab:balanced_accuracy_LLM}, DeepSeek R1, GPT-4o, and GPT-o1 all achieve balanced accuracy above 99.5\% across S1–S4. DeepSeek R1 reaches 100\% in S1 and remains highly accurate in others (99.5\%–99.9\%), while GPT-4o and GPT-o1 show only minor variations. These results indicate that the choice of LLM introduces negligible differences, allowing practitioners to flexibly select models based on availability, cost, or deployment requirements.
\begin{table}[htbp]
\centering
\caption{Balanced Accuracy of Different LLMs.}
\setlength{\tabcolsep}{0.5em}
\begin{tabular}{c|rrr}
\hline
\multicolumn{1}{c|}{\multirow{2}{*}{\textbf{Attack Scenario}}} & \multicolumn{3}{c}{\textbf{LLMs}} \\
\cline{2-4}
 & \textbf{DeepSeek-R1} & \textbf{GPT-4o} & \textbf{GPT-o1} \\
\hline\hline
S1 & 100.0\% & 100.0\% & 100.0\% \\
S2 & 99.9\% & 99.9\%& 99.9\% \\
S3 & 99.9\% & 99.9\% & 99.7\% \\
S4 & 99.5\% & 99.5\% & 99.5\% \\
\hline
\end{tabular}
\label{tab:balanced_accuracy_LLM}
\end{table}

\noindent
\textbf{Impact of Additional Knowledge Units.}
We test robustness by randomly adding 5–30 Linux attack knowledge units to the knowledge base. As reported in \Cref{tab:balanced_accuracy_units}, performance remains consistently high, with  Balanced Accuracy stable at 99.5\%–100\% across S1–S4. This shows that \tool is insensitive to the presence of extra knowledge units.
\begin{table}[htbp]
\centering
\caption{Balanced Accuracy with Increasing Units.}
\small
\setlength{\tabcolsep}{0.1em}
\begin{tabular}{c|rrrrrr}
\hline
\multicolumn{1}{c|}{\multirow{1}{*}{\textbf{Attack}}} & \multicolumn{6}{c}{\textbf{Additional Knowledge Units}} \\
\cline{2-7}
\multicolumn{1}{c|}{\multirow{1}{*}{\textbf{Scenario}}} &  \multicolumn{1}{c}{\multirow{1}{*}{\textbf{5}}}  & \multicolumn{1}{c}{\multirow{1}{*}{\textbf{10}}} & \multicolumn{1}{c}{\multirow{1}{*}{\textbf{15}}} & \multicolumn{1}{c}{\multirow{1}{*}{\textbf{20}}} & \multicolumn{1}{c}{\multirow{1}{*}{\textbf{25}}} & \multicolumn{1}{c}{\multirow{1}{*}{\textbf{30}}} \\
\hline\hline
S1 & 100.0\% & 100.0\% & 100.0\% & 100.0\% & 100.0\% & 100.00\% \\
S2 & 99.9\% & 99.9\% & 99.9\% & 99.9\% & 99.9\% & 99.9\% \\
S3 & 99.9\% & 99.9\% & 99.9\% & 99.9\% & 99.9\% & 99.9\% \\
S4 &  99.5\% & 99.5\% & 99.5\% & 99.5\% & 99.5\% & 99.5\% \\
\hline
\end{tabular}
\label{tab:balanced_accuracy_units}
\end{table}

\noindent
\textbf{Impact of Similarity Metrics.}
Finally, we evaluate three similarity metrics used for retrieval in the RAG system: cosine similarity, inner product, and euclidean distance. As shown in \Cref{tab:balanced_accuracy_similarity}, all metrics yield nearly identical performance, with Balanced Accuracy consistently above 99.9\%. This demonstrates that \tool is robust to the choice of similarity function, as retrieval remains stable across metrics.
\begin{table}[htbp]
\centering
\caption{Balanced Accuracy of Different Similarity Metrics.}
\setlength{\tabcolsep}{0.2em}
\begin{tabular}{c|ccc}
\hline
\multicolumn{1}{c|}{\multirow{2}{*}{\textbf{Attack Scenario}}} & \multicolumn{3}{c}{\textbf{Similarity Metrics}} \\
\cline{2-4}
 & \textbf{Cosine} & \textbf{Inner Product} & \textbf{Euclidean} \\
\hline\hline
S1 & 100.0\% & 100.0\% & 100.0\% \\
S2 & 99.9\% & 99.9\%& 99.9\% \\
S3 & 99.9\% & 99.9\% & 99.9\% \\
S4 & 99.5\% & 99.5\% & 99.5\% \\
\hline
\end{tabular}
\label{tab:balanced_accuracy_similarity}
\end{table}

\subsection{Case Study}
\label{sec:case study}

To illustrate the effectiveness of \tool in automated forensic analysis, we present a case study based on the EB-P1 scenario. This example demonstrates how \tool helps analysts uncover complex attack behaviors with minimal manual intervention. Since the original attack subgraphs are too large and complex to display, we simplify the visualization by retaining only the critical attack steps and omitting auxiliary events for clarity.

\begin{figure}[h!t]
  \centering
  \includegraphics[width=0.48\textwidth]{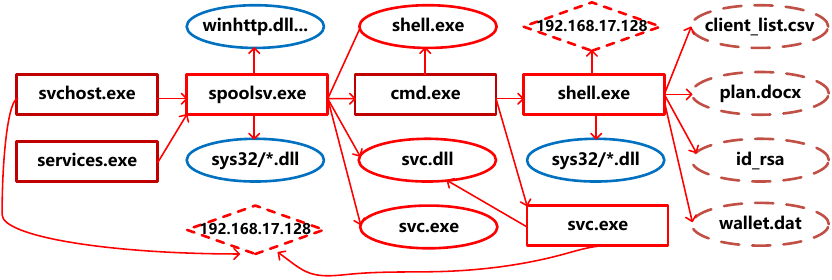}
  \caption{Provenance summary graph of the EB-P1 attack scenario, reconstructed automatically by \tool. Red nodes and edges denote attack-related activities.}
  \label{fig:case_study}
\end{figure}

As shown in Figure~\ref{fig:case_study}, the attack exploits the well-known SMBv1 vulnerability (CVE-2017-0144), which enables remote code execution via specially crafted SMB packets. This flaw, affecting a core Windows protocol for file and printer sharing, allows attackers to trigger memory corruption and execute arbitrary shellcode in kernel space.

\noindent
\ding{182} The attack begins with the compromise of a Windows host at \texttt{192.168.17.128} through a malicious SMB transaction.

\noindent
\ding{183} The injected shellcode is mapped into the memory space of the legitimate \texttt{spoolsv.exe}, enabling stealthy control.

\noindent
\ding{184} From this foothold, the adversary executes \texttt{shell.exe} to deploy additional payloads and run system-level commands. To establish persistence, a malicious service \texttt{svc.exe} is installed and registered with the Windows Service Control Manager, ensuring automatic execution on reboot.

\noindent
\ding{185} The forensic trace reveals unauthorized access to sensitive assets, including SSH private keys (\texttt{id\_rsa}), internal documents (\texttt{plan.docx}), and customer records (\texttt{client\_list.csv}). Using provenance~graph~expansion, \tool reconstructs the full attack chain: initial SMB exploitation, shellcode injection into a trusted process, persistence via service registration, and data exfiltration.
\section{Discussion}\label{sec:Discussion}

\subsection{Unknown Threats}
Like other approaches that rely on annotated attack data, our method requires labeled knowledge to populate the database. If adversaries adopt tactics that are entirely absent from existing knowledge, investigation may fail. However, unlike small models that require costly retraining and risk catastrophic forgetting, \tool adapts by simply appending new units to its knowledge base. This design supports rapid updates without degrading prior capabilities. In practice, curated threat intelligence from research and industry can be leveraged to continuously evolve and strengthen the knowledge base.

\subsection{Against False Alerts}
Our work focuses on attack investigation based on security alerts, while the detection or reduction of false alerts is beyond our scope.
Recent approaches \cite{relatedwork_rule_graph_based:dong2023distdet,kairos:cheng2024kairos} have proposed to reduce false alerts in detection, and \tool can work with these approaches to achieve better investigation capability. Meanwhile, investigation reports generated from false alerts fail to form complete attack chains, making them identifiable and thus discarded.

\subsection{Future Work}
Our current knowledge base follows the Cyber Kill Chain model, which provides a high-level abstraction of multi-phase attacks by dividing behaviors into seven sequential phases. While effective, this structure may overlook the fine-grained techniques observed in real-world intrusions. By contrast, the MITRE ATT\&CK framework~\cite{mitre_ttps} offers a more detailed taxonomy of adversarial techniques organized by tactical goals. Incorporating ATT\&CK into our design could significantly improve forensic precision and the quality of generated investigation reports, but would require more sophisticated organization of knowledge units to manage the added complexity. As future work, we plan to explore ATT\&CK-guided knowledge base structuring to enhance both expressiveness and adaptability of \tool.

\section{Related Work}\label{sec:Related}
\tool is closely related to prior efforts on provenance-based attack investigation, which can be broadly categorized into heuristics-based and learning-based approaches.

\smallskip
\noindent\textbf{Heuristics-based provenance analysis.}
Heuristics-based approaches~\cite{relatedwork_rule_based:hossain2017sleuth,relatedwork_static_based:hassan2019nodoze,relatedwork_rule_based:milajerdi2019holmes,relatedwork_rule:milajerdi2019poirot,relatedwork_rule_based:fang2022back,xu2022depcomm} mitigate dependency explosion by applying handcrafted rules tailored to known attack patterns. For example, SLEUTH~\cite{relatedwork_rule_based:hossain2017sleuth} propagates trust labels across provenance graphs to trace attacks in real time. NoDoze~\cite{relatedwork_static_based:hassan2019nodoze} ranks alerts through causal tracing and anomaly scoring based on historical event frequencies, prioritizing suspicious alerts. Poirot~\cite{relatedwork_rule:milajerdi2019poirot} detects malicious activities using predefined behavioral patterns, while Fang et al.\cite{relatedwork_rule_based:fang2022back} assign edge weights and back-propagate impact to filter irrelevant dependencies. DEPCOMM\cite{xu2022depcomm} compresses dependency graphs into concise process-centric communities, and HOLMES~\cite{relatedwork_rule_based:milajerdi2019holmes} correlates low-level audit logs with high-level adversarial TTPs. RapSheet~\cite{relatedwork_rule_based:RapSheet:hassan2020tactical} extends detector-generated alerts into tactical provenance graphs for triage.
While effective in reducing noise, these methods rely heavily on manually crafted rules, which are costly to construct, difficult to update, and limited in capturing novel or evolving adversarial behaviors. In contrast, \tool leverages LLMs augmented with structured threat knowledge to reason about attacks in a human-like manner, enabling adaptive and comprehensive investigation.

\smallskip
\noindent\textbf{Learning-based provenance analysis.}
Learning-based approaches~\cite{airtag_2023,atlas_2021,related_learning:kapoor2021prov,related_learning:liu2019log2vec,related_learning:li2021hierarchical,related_learning:pei2016hercule,related_learning:wang2022threatrace,kairos:cheng2024kairos,jiang2025orthrus} train models over provenance graphs or event sequences to capture benign and malicious behaviors. Log2Vec~\cite{related_learning:liu2019log2vec} embeds provenance graphs into low-dimensional vectors for clustering, while Prov-gem~\cite{related_learning:kapoor2021prov} applies relation-aware embeddings with self-attention for classification. Hercule~\cite{related_learning:pei2016hercule} uses community discovery on correlated logs to reconstruct attack stories. More recent systems adopt graph neural networks (GNNs): Kairos~\cite{kairos:cheng2024kairos} applies a temporal GNN encoder–decoder for anomaly detection in streaming logs, and ORTHRUS~\cite{jiang2025orthrus} integrates GNN-based detection with dependency-guided reconstruction to improve attribution quality. \atlas~\cite{atlas_2021} trains a sequence model on an abstracted vocabulary of provenance events, while \airtag~\cite{airtag_2023} uses unsupervised modeling for binary classification of logs.
Although these methods advance automation, they are constrained by small model capacity, limited generalization across platforms, and frequent reliance on platform-specific feature engineering, requiring retraining when environments change. By contrast, \tool employs LLMs with knowledge retrieved from a Kill Chain aligned knowledge base, enabling adaptation to heterogeneous logs and evolving attack tactics without retraining or feature engineering.

\section{Conclusion}\label{sec:Conclusion}
We presented \tool, the first LLM-empowered attack investigation framework augmented with a dynamically updatable Kill Chain–aligned knowledge base. By integrating provenance-graph exploration, iterative causal reasoning, and retrieval-augmented knowledge guidance, \tool reconstructs multi-phase intrusion scenarios from massive heterogeneous logs and generates coherent analyst-ready reports without retraining or platform-specific tuning. Evaluation across 15 attack scenarios spanning single-host, multi-host, Windows, Linux, and stealthy shellcode injection shows that \tool achieves highly accurate forensic analysis with extremely low false positives, significantly outperforming \atlas while generalizing across platforms, attack types, and alert entry points. These results demonstrate that \tool provides an effective and scalable foundation for automated attack investigation, advancing real-world defense through accurate, adaptive, and knowledge-driven analysis.

\clearpage

\section*{Ethical Considerations}
This work focuses on developing a framework for automated attack investigation using host logs and a knowledge-guided reasoning process. Our study does not involve human subjects, personal data, or any experiments on production systems. All logs used in evaluation are either synthetically generated or drawn from publicly available security benchmarks, which do not contain sensitive personal information. When discussing advanced persistent threats (APTs) and attack techniques, we rely on threat intelligence reports and open datasets rather than privately collected incident data.

\section*{Open Science}
\label{sec:code}
\textbf{Datasets Availability.}
The publicly released datasets used in our evaluation are available from their original sources~\cite{atlas_2021,airtag_2023}. 
In addition, the simulated attack datasets generated for our experiments are available at \url{https://anonymous.4open.science/r/Themis-LLM-B4EC}.

\noindent\textbf{Software Artifacts Availability.} The source code is publicly available at \url{https://anonymous.4open.science/r/Themis-LLM-B4EC}. 
We also provide detailed instructions on how to reproduce the experimental results.

\bibliographystyle{plain}
\bibliography{refs}

\appendix
\begin{center}
   \LARGE \textbf{Appendix} 
\end{center}
\section{Prompt Templates}
\subsection{Prompt for Kill-Chain Annotation}
\label{llm:prompt:annotate}
\begin{tcolorbox}[title=Prompt for Kill-Chain Annotation,breakable, before skip=0pt, after skip=0pt]
You are an expert in Kill Chain knowledge, write the analysis. Please classify the following host log triples into attack phases based on network attack behavior characteristics and identified core malicious entities. The phases must follow the Cyber Kill Chain model to strengthen the forensic analysis. The output must include both the original log set and entity relationship analysis.

Contexts:[malicious\_entities],[sequences]

Task: 
1. Identify critical stages of the attack lifecycle (e.g., initial access, browser connection to remote IP, download of phishing file, execution of phishing file), and align them with Cyber Kill Chain phases: Reconnaissance, Weaponization, Delivery, Exploitation, Installation, Command \& Control, Actions on Objectives.

2. Detect malicious file deployment patterns

3. Infer causality between the current and previous window based on behavioral continuity...

4. Analyze invocation relationships between entities.

5. Explicitly include the names of the relevant entities.

6. ...

Processing Rules: 
1. Strictly preserve the original temporal order.

2. Merge operations with the same attack intent into a single phase.

3. Separate different attack vectors into independent phases.

4. Merge multiple actions triggered by the same entity.

5.evidence\_set must fully retain the original log entries.

6. Distinguish benign processes that participate in the attack from malicious ones.

7. ...

Output Format:
Return the result as a JSON array...

\end{tcolorbox}

\subsection{Prompt for Causal Reasoning}
\label{llm:prompt:reasoning}
\begin{tcolorbox}[title=Prompt for Causal Reasoning,breakable]
You are a cyber forensic analyst investigating potential attacks in preprocessed system logs...

Given a sequence of attack-related behavior records...

The logs are structured as (subject, action, object) triple...

Contexts:

Known malicious entities from alerts:[payload]

Previously inferred suspicious entities: [detected]

Current log sequence : [sequence]

Prior reasoning summary: [summary]

attack pattern from knowledge base:[augmentation knowledge ]

Task: 

1. Analyze the current log sequence in temporal order ...

2. Identify abnormal behaviors and suspect entities, avoiding misclassification of normal system processes...

3.Infer causality between the current and previous window based on behavioral continuity...

4...

Answer:

Output Format:
Return the result as a JSON array...
\end{tcolorbox}
\subsection{Prompt for Attack Report Generation}
\label{llm:prompt:generation}
\begin{tcolorbox}[title=Prompt for Causal Reasoning,breakable, before skip=0pt, after skip=0pt]
You are a cyber forensic analyst responsible for consolidating reasoning results into a comprehensive attack investigation report. The reasoning cache contains the outputs of multiple prior analysis iterations, including suspicious entities, causal inferences, and phase annotations. Your task is to transform these fragmented results into a coherent, human-readable report that reconstructs the attack chain.

Contexts:
Aggregated reasoning cache: [reasoning\_cache]
Detected suspicious entities across iterations: [detected]
Task: 
1. Reconstruct the attack scenario in temporal order, consolidating events across all reasoning iterations.

2. Organize the scenario into Kill Chain phases (e.g., Initial Access, Persistence, Lateral Movement, Exfiltration).

3.Summarize abnormal behaviors and their causal relationships, highlighting how each suspicious entity contributed to the attack progression.

4. Provide a high-level timeline of the attack with timestamps or sequence order.

5. ...

Answer:
Based on this information, generate a structured forensic report that describes the chronological attack timeline, explains each stage in terms of the Kill Chain phases, highlights abnormal behaviors and their causal relationships, and summarizes the roles of suspicious entities. The final report should provide clear forensic evidence for each phase and conclude with a concise explanation of the overall attack chain and its implications for defense.
\end{tcolorbox}

\end{document}